\documentclass[aps,prb,twocolumn,showpacs,10pt]{revtex4-1}
\usepackage{graphicx}
\usepackage{amssymb}
\usepackage{amsmath,bm}
\usepackage{comment}
\usepackage{footmisc}
\usepackage{color}
\usepackage{epsfig}
\usepackage{graphicx,amsmath,amssymb,amsfonts,sidecap,color,hyperref}
\usepackage{subfigure}
\newcommand{\nn}{\nonumber\\}

\begin{document}

\title{ Majorana Kramers pairs in  Rashba double nanowires with interactions and disorder}

\author{Manisha Thakurathi$^{1}$}
\author{Pascal Simon$^{2}$}
\author{Ipsita Mandal$^{1,3}$}
\author{Jelena Klinovaja$^{1}$}
\author{Daniel Loss$^{1}$}
\affiliation{$^{1}$Department of Physics, University of Basel, Klingelbergstrasse 82, CH-4056 Basel, Switzerland}
\affiliation{$^{2}$Laboratoire de Physique des Solides, CNRS UMR-8502, Universit\'e Paris Sud, 91405 Orsay Cedex, France}
\affiliation{$^{3}$Max-Planck Institute for the Physics of Complex Systems, Noethnitzer Str.  38, 01187, Dresden, Germany}

\date{\today}

\begin{abstract}
We analyze the effects of electron-electron interactions and disorder on a Rashba double-nanowire setup coupled to an $s$-wave superconductor, which has been recently proposed as a versatile platform to generate Kramers pairs of Majorana bound states in the absence of magnetic fields. We identify the regime of parameters for which these Kramers pairs are stable against interaction and disorder effects. We use bosonization, perturbative renormalization group, and replica techniques to derive the flow equations for various parameters of the model and evaluate the corresponding phase diagram with topological and disorder-dominated phases. We confirm aforementioned results by considering a more microscopic approach which starts from the tunneling Hamiltonian between the three-dimensional $s$-wave superconductor and the nanowires. We find again that the interaction drives the system into the topological phase and, as the strength of the source term coming from the tunneling Hamiltonian increases, strong electron-electron interactions are required to reach the topological phase. 
\end{abstract}

\maketitle

\section {Introduction}
Over the past decade or so many studies on topological phases in condensed matter systems have been performed \cite{Hasan}.
In particular, Majorana bound states (MBSs) in such systems have attracted a lot of attention  because of their potential application in topological quantum computation based on their non-Abelian braiding statistics \cite{Kitaev}. There have been many advancements both  theoretically  as well as experimentally on MBSs in semiconductor nanowires (NWs) with proximity gap and Rashba spin orbit interaction and on their detection~\cite{Mourik,Das,deng2012anomalous,churchill2013superconductor,albrecht2016exponential,
oreg2010helical,sticlet2012spin,dominguez2012dynamical,prada2012transport,maier2014majorana,sarma2012splitting,leijnse2012introduction,
klinovaja2012transition,bjornson2013vortex,weithofer2014electron,dmytruk2015cavity,
JK1,Denis1,Alicea,Lutchyn,Schrade2,Denis2,MT1,Olesia,Kouwenhoven,Prada,Marcus,Ptok}.
So far, most of the studies on MBSs have been focused on the generation of these exotic states in the presence of magnetic fields. However, recently it has been shown that MBSs can also be generated in the absence of magnetic fields \cite{Nagaosa,Zhang,Tewari,Keselman1,Keselman2,Schrade1,JK3, JK4,JK5,Flensberg,Grifoni}, having the advantage to avoid detrimental effects of the magnetic field on the host $s$-wave superconductor which is needed to induce proximity gaps in the NWs. The resulting twofold degeneracy of these MBSs is protected by time-reversal symmetry and therefore gives rise to  Kramers pairs of MBSs (KMBSs).

The key property of MBSs is their robustness against local perturbations. Therefore for the low-energy physics, it becomes crucial to consider the effects of electron-electron interactions\cite{Fidkowsky,Turner,Suhas,Stoudenmire2011,Sela2011,Lutchyn2011,Hassler2012,Kells2015,Miao2017,Dominguez} and disorder \cite{Motrunich2001,Brouwer2011,Akhmerov2011,Stanescu2011,Brouwer2011_2,Stanescu2011_2,Gottardi2013,Rainis,MT2}, as these perturbations, taken independently, are able to affect the topological protection of the MBSs  even if they preserve time-reversal symmetry. 
When both disorder and electron-electron interactions  are taken into account, a perturbative treatment in disorder and pairing indicates that they indeed reinforce each other to destroy the topological gap \cite{Lobos2012}, a result  corroborated by a Gaussian variational ansatz \cite{Crepin2014} and further extended  in the opposite strong disorder limit using the density matrix renormalization group approach \cite{Gergs}, although some recent investigations indicate that in the moderate disorder regime both effects can cooperate to actually stabilize and even enhance the topological order \cite{Gergs,Kells2017}.

In the present work, we consider a time-reversal invariant system, 
which supports KMBSs in the topological phase, and analyze the stability of this phase against bulk disorder and electron-electron interactions, using bosonization and Luttinger liquid (LL) techniques.  
In general, if  we start from a gapped superconducting phase and switch on electron-electron interactions and/or disorder, their bulk effect is qualitatively not essential as long as their corresponding strengths are smaller than the effective gap in the system. Another approach is to start from a gapless phase and treat both the proximity effect and electron-electron interactions and/or disorder on equal footing by treating them as perturbations and determining which ones dominate\cite{Suhas,Lobos2012,Crepin2014}.
Here we follow the latter approach, starting from a gapless phase, and analyze the competition between  proximity, interaction, and disorder  effects using a perturbative renormalization group (RG) approach. 

\begin{figure}[t]
\includegraphics[width=0.8\linewidth]{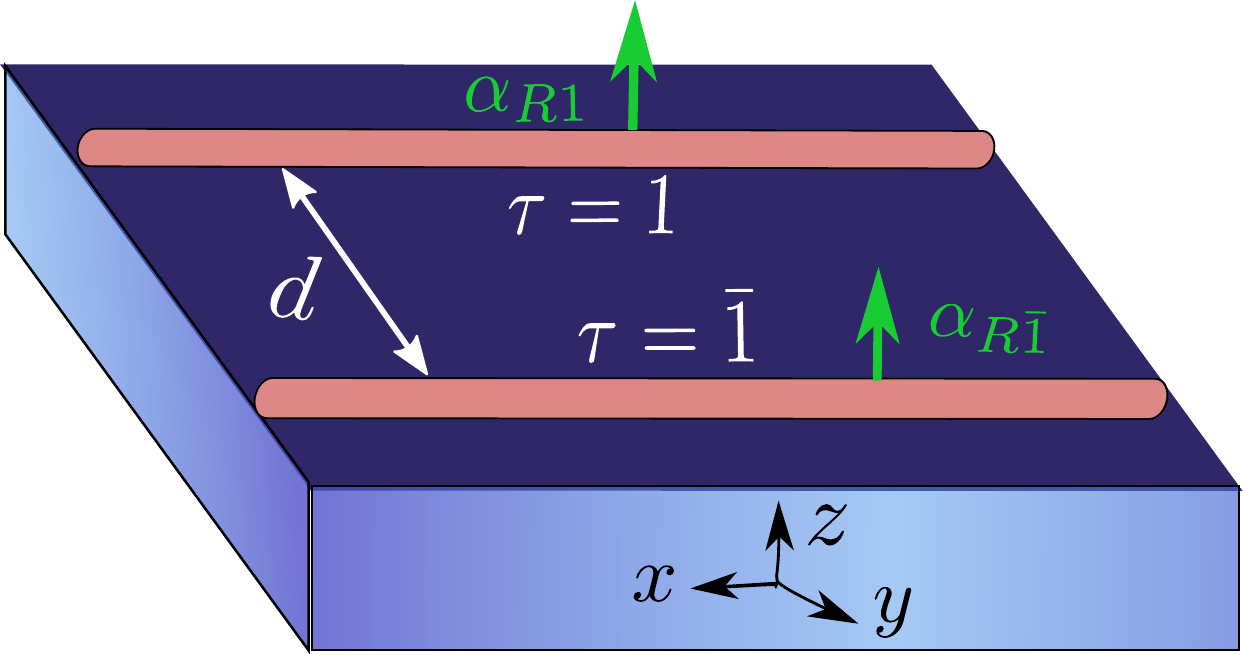}
\caption{Sketch of the double-NW setup consisting of  two Rashba NWs (brown strips) of length $L$, which are aligned along $x$-direction, separated by a distance $d$, and tunnel coupled to a three-dimensional $s$-wave superconductor (blue slab). The NWs are labeled by the index $\tau =1$ $(\bar 1)$ referring to the upper (lower) NW. The direction of the SOI vectors $\mathbf{\alpha_{R\tau}}$ in both NWs is chosen along $z$-direction.}
\label{fig00}
\end{figure}

As a model system we consider the double-NW setup proposed in Ref. [\onlinecite{JK3}], depicted in Fig. \ref{fig00}, which consists of two one-dimensional NWs labeled by $\tau=1 \,(\bar 1)$ for the upper (lower) NW, with Rashba spin orbit interaction (SOI). The NWs are in proximity to an $s$-wave superconductor underneath. This  geometry gives rise to two classes of proximity induced pairing terms, the first one is intrawire  pairing due to tunneling of Cooper pairs as a whole to either of the NWs. The second class is the interwire  pairing corresponding to  crossed Andreev reflection \cite{Feinberg,Schoenenberger,Heiblum,Tarucha,Recher_Sukhorukov,DL1,Bena}. It has been shown that this setup  can support two MBSs at each end of the double-NW setup which are time-reversal partners of each other, corresponding to KMBSs, provided the interwire crossed Andreev pairing gap
exceeds the  intrawire pairing gap. However, in the non-interacting system, it has been established that the value of the interwire pairing gap is always smaller than the intrawire one \cite{Reeg_a}. The goal of this work is to show that interactions can reverse the situation and enable the system
to become topological.

We approach this problem in two different ways using renormalization group (RG) techniques. First, we start from an effective model where the superconducting pairing amplitudes in the NWs are introduced as model parameters and analyze the behavior of these terms in the presence of electron-electron interaction and disorder. We find a physically relevant regime 
where the interwire crossed Andreev pairing amplitude exceeds the intrawire one due to interactions, and thus the NW system can reach the topological phase and host a KMBS at each end of the setup. In particular, this topological regime is reached when the repulsive interaction, characterized by charge and spin LL parameters $K_{\tau c}$ and $K_{\tau s}$ for each NW $\tau$, satisfy $K_{\tau c}<1$ and $K_{\tau s}\geq 1$.

Moreover, we determine the full phase diagram as a function of interaction and disorder strengths and explore a wide range of parameter values for which the system can be topological and  host KMBSs. 
In a second, more microscopic approach, we start from a model which includes the tunnel coupling between NWs and superconductor and thereby  the superconducting gaps in NWs are generated due to tunneling of Cooper pairs from the superconductor into the NWs in the simultaneous presence
of electron-electron interactions. We derive and analyze the RG flow equations here as well, which now contain a `source term' (coming from the tunneling Hamiltonian) that flows under renormalization and thereby generates the pairing terms. Such source terms have been considered before  in the study of
proximity gaps in topological insulators\cite{Recher}.
Again, we find that the repulsive interactions can drive the system from the trivial to the topological phase,  however, the required strength of the electron-electron interactions to reach the topological phase 
must be larger compared to the effective pairing model.

The outline of the paper is as follows. In Sec. \ref{model}, we introduce the  model for the double-NW system. In Sec. \ref{bos}, we apply bosonization techniques to include electron-electron interactions. We briefly review the replica method for the treatment of disorder averaging in the Sec. \ref{dis}, followed by the RG analysis in Sec. \ref{rg}. In Sec. \ref{rgt}, we introduce the tunneling Hamiltonian between NWs and superconductor, calculate the RG flow equations from the source terms, and confirm the results obtained in Sec. \ref{rg}. Finally, in Sec. \ref{con} we conclude with a summary and outlook. Technical details are deferred to Appendixes \ref{alp}-\ref {argt}. 

\section{Model} \label{model}
We consider a double-NW setup, depicted in Fig. \ref{fig00}, which consists of two Rashba NWs (of length $L$) labeled by an index $\tau=1$ ($\tau=\bar1$) for upper (lower) NW ~\cite{JK3}. The two NWs aligned along the $x$-direction are in the proximity of an $s$-wave superconductor.
The Hamiltonian of the non-interacting and disorder-free system has the form
\begin{align}
H=H_0+H_{sc},
\end{align}
where $H_0$ describes the kinetic part of the Hamiltonian as $H_{sc}$ describes the superconducting pairing arising in the NWs due to the coupling to a bulk SC.
The kinetic part is defined as
\begin{align}
&H_0 =\sum_\tau \int dx\ \Big[ \sum_{\sigma} \Psi_{\tau\sigma}^\dagger(x) \left( \frac{-\hbar^2 \partial_x^2}{2m}  - \mu_\tau \right)\Psi_{\tau\sigma}(x) \nn
&\hspace{70pt}- i \sum_{\sigma,\sigma'} \alpha_{R\tau}  \Psi_{\tau\sigma}^\dagger (\sigma_3)_{\sigma\sigma'} \partial_x\Psi_{\tau\sigma'}\Big],
\end{align}
where $\mu_\tau, \alpha_{R \tau}>0$ are the chemical potential and Rashba SOI strength in the $\tau$-NW, respectively. Here, the operator $\Psi^\dagger_{\tau\sigma}(x)$ [$\Psi_{\tau\sigma}(x)$] creates (annihilates) an electron of band mass $m$ with  spin $\sigma/2 =\pm 1/2$ at position $x$ of the $\tau$-NW. The Pauli matrices $\sigma_{1,2,3}$ act on the spin of the electron. In both NWs,  the SOI vectors are aligned in the $z$-direction. The energy spectrum for electrons with spin component $\sigma$ in the $\tau$-NW  is given by 
\begin{equation}
E_{\tau\sigma} = \hbar^2 (k-\sigma k_{so,\tau})^2/2m,
\end{equation}
where $k_{so,\tau}=m\,\alpha_{R\tau}/\hbar^2$ is the SOI wavevector with $E_{so,\tau}=\hbar^2k_{so,\tau}^2/2m$ being the SOI energy. For simplicity, we tune the chemical potentials in both NWs to the corresponding SOI energy, $\mu_\tau = E_{so,\tau}$.

The second term in the Hamiltonian $H$ is the proximity-induced superconducting pairing term, $H_{sc}$, and has two contributions corresponding to intrawire ($H_{s}$) and interwire ($H_{c}$) pairings ~\cite{JK3}.  The intrawire pairing  of strength $\Delta_{\tau}$ accounts for the tunneling of Cooper pairs as a whole from the superconductor to the $\tau$-NW. However, when the electrons from the same Cooper pair separate and each electron tunnels into a different NW, this gives rise to the interwire pairing gap of strength $\Delta_{c}$. This process is referred to as crossed-Andreev pairing and has been investigated in detail for the double-NW setup considered here \cite{Reeg_a} but in the absence of electron-electron interactions and for disorder-free NWs. The corresponding pairing terms in the Hamiltonian are written as
\begin{align}
H_{sc}=&H_{s}+H_{c}\nn
= &\sum_{\tau,\sigma,\sigma'} \int dx\,  \Big[ \frac{\Delta_\tau}{2}
 \Psi_{\tau\sigma} \, (i \, \sigma_2)_{\sigma\sigma'}\, \Psi_{\tau\sigma'} \nn
&\hspace*{1.7cm}+\frac{\Delta_{c}}{2} 
 \Psi_{\tau\sigma}\, (i \, \sigma_2)_{\sigma\sigma'} \, \Psi_{\bar \tau\sigma'}+  \text{H.c.}\,\Big],
\label{HSC}
\end{align}
where ${\bar \tau}=-\tau$. The gap at $k=0$ in the spectrum of the double-NW setup is given by $\Delta_g=\sqrt{|\Delta_c^2-\Delta_1\Delta_{\bar 1}|}$ \cite{JK3}. As a result, the topological phase hosting two MBSs at each end of the double-NW setup is defined by
\begin{equation}
\Delta_c^2>\Delta_1\Delta_{\bar 1}.
\label{topocrit}
\end{equation} 
The topological criterion given by Eq. (\ref{topocrit}) cannot be satisfied for non-interacting systems\cite{Reeg_a,Schrade2} (unless a magnetic field is turned on which breaks time-reversal symmetry \cite{Haim}). However, in the presence of interactions it has been argued~\cite{JK3} that the crossed Andreev process is favored over the direct one as the latter one is relatively stronger suppressed by electron-electron interactions when
the two electrons of the same Cooper pair enter the same NW, meaning that we add simultaneously two charges, while we add only one charge per NW in the crossed Andreev process.  Similar arguments underly the mechanism of Cooper pair splitters based on quantum dots~\cite{Recher_Sukhorukov} or NWs~\cite{DL1,Bena} where also crossed Andreev processes get favored over direct ones by
interaction effects. Such effects are experimentally well established for transport currents through setups similar to the one shown in Fig.~\ref{fig00} but where the NWs are replaced by quantum dots~\cite{Schoenenberger, Heiblum,Tarucha}.
In the following we wish to study the proximity effect in the presence of electron-electron interactions in the NWs and show that Eq. (\ref{topocrit}) can indeed be satisfied
under certain conditions.
For this we have to treat the interaction effects in the one-dimensional NWs non-perturbatively, making use of bosonization, Luttinger liquid, and renormalization group techniques, as described in the following sections.

\begin{figure}[t]
\includegraphics[width=0.6\linewidth]{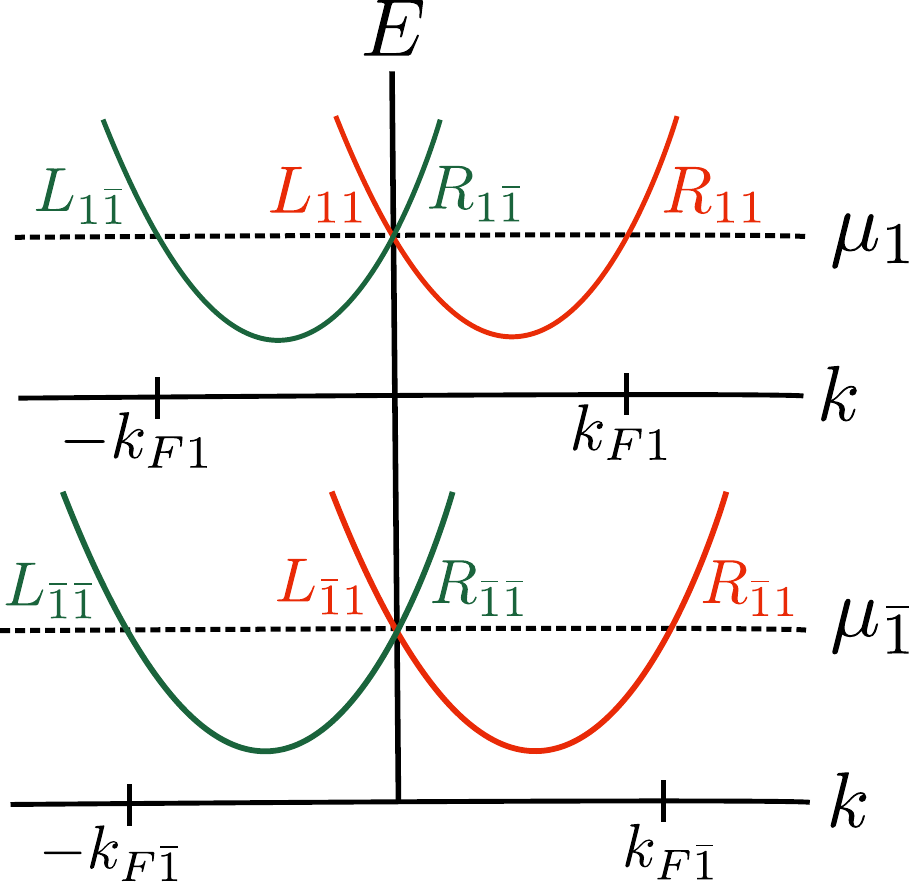}
\caption{Spectrum of two spatially separated Rashba NWs labeled by $\tau=1$ ($\tau=\bar 1$) for upper (lower) NW, with different Rashba SOI momenta $k_{k_F\tau}$. The red (green) color code is for electron spin, $\sigma/2 = +1/2$ ($\sigma/2 = -1/2$). The chemical potential ($\mu_\tau $) in both NWs is tuned to the crossing point between $\sigma/2 = +1/2$ and $\sigma/2 = -1/2$  electrons at $k=0$. We linearize the spectrum around the Fermi points $k_{F\tau}$ and $k=0$ and label the slowly moving right (left) electron fields  as $R_{\tau \sigma}$ ($L_{\tau \sigma}$). }
\label{spectrumbar}
\end{figure}

\section{Bosonization} \label{bos}

In this section, we first linearize the spectrum around the Fermi points $k=0$ and $k=\pm k_{F\tau}\equiv \pm 2k_{so,\tau}$ (see Fig.~\ref{spectrumbar}) and subsequently bosonize the Hamiltonian to include the electron-electron interactions. We first decompose the Fermi fields in their right and left movers \cite{JK2},
\begin{align}
&\Psi_{11}=R_{11}e^{i k_{F1}  x}+L_{11} \,,\nn
&\Psi_{1\bar 1}= R_{1\bar1}+ L_{1\bar1}e^{-ik_{F 1}x} \,,\nn
&\Psi_{\bar 1 1}=R_{\bar 11} e^{ik_{F \bar1}x} + L_{\bar 11} \,,\nn
&\Psi_{\bar1\bar 1}=R_{\bar 1\bar1} +L_{\bar 1\bar1} e^{-ik_{F \bar 1}x}\,,
\end{align}
where $R_{\tau\sigma}(x)$ [$L_{\tau\sigma}(x)$] is the slowly-varying right(left)-moving  field of an electron at position $x$  in the $\tau$-NW with spin $\sigma$. The kinetic energy and SOI Hamiltonian $H_0$ [see Eq. (\ref{H0})] reduces to
\begin{align}
&H_0= \sum_{\tau,\sigma}
i\, \hbar \, \upsilon_{F\tau} \int dx\,  [ \, L_{\tau\sigma}^\dagger\partial_xL_{\tau\sigma} - R_{\tau\sigma}^\dagger\partial_xR_{\tau\sigma} \, ],
\label{H0}
\end{align}
where $\upsilon_{F\tau} = \hbar k_{F \tau}/m$ is the Fermi velocity for $\tau$-NW. We note that the interwire pairing term $\Delta_{c}$ acts only on momenta close to zero, as described in Ref. [\onlinecite{JK3}]. We also divide the intrawire term into two parts, one ($\Delta^{ext}_{\tau}$) term acts on states with momenta close to $ k_{F \tau} $ (exterior branches)  while the other term ($\Delta^{int}_{\tau}$) acts on states with momenta close to zero (interior branches).  The intrawire and interwire proximity-induced pairing terms can then be rewritten as 
\begin{align}
&H_{s}= \sum_{\tau}  \frac{\Delta^{ext}_{\tau}}{2}
\int dx\,  \left( R_{\tau1}^\dagger L_{\tau\bar1}^\dagger
- L_{\tau\bar1}^\dagger R_{\tau1}^\dagger \right )
\nonumber\\
&\hspace{20pt}
+ \sum_{\tau}  \frac{\Delta^{ int}_{\tau}}{2}
\int dx\,  \left( L_{\tau 1}^\dagger R_{\tau\bar1}^\dagger- R_{\tau\bar1}^\dagger L_{\tau1}^\dagger  \right) +  \text{H.c.}\,, 
\label{HSC1}\\
&H_{c}=\frac{\Delta_{c}}{2} \int dx\, \Big( L_{\bar 11}^\dagger R_{ 1\bar1}^\dagger - R_{ 1\bar1}^\dagger L_{\bar 11}^\dagger \nonumber\\
& \hspace{2cm} +L_{11}^\dagger R_{ \bar1\bar1}^\dagger -  R_{\bar 1\bar1}^\dagger L_{ 11}^\dagger 
\Big)+ \text{H.c.}
\label{HSC2}
\end{align}
 Furthermore, we perform the standard bosonization of fermions by introducing the charge ($\phi_{\tau c}$, $\theta_{\tau c}$) and spin ($\phi_{\tau s}$, $\theta_{\tau s}$) bosonic fields \cite{TG}. These fields obey the bosonic commutation relations $[\phi_{\tau,c/s}(x),\theta_{\tau',c/s}(x')]= i \pi \delta_{\tau \tau'}\, \text{sgn}(x'-x)/2 $. We write the left and right moving fermions in terms of the charge-spin bosonic fields as  
\begin{align}
R_{\tau\sigma}&=\frac{1}{\sqrt {2 \pi \alpha}}  e^{-\frac{i}{\sqrt 2}[ \phi_{\tau,c}-\theta_{\tau,c}+\sigma \left (\phi_{\tau,s}-\theta_{\tau,s} \right ) ]}\,,\nn
L_{\tau\sigma}&=\frac{1} {\sqrt {2 \pi \alpha} } e^{\frac{i}{\sqrt 2}[ \phi_{\tau,c}+\theta_{\tau,c}+\sigma \left (\phi_{\tau,s}+\theta_{\tau,s} \right ) ]}\,,
\label{chiralfield}
\end{align}
where $\alpha$ is the ultraviolet (short-distance) cutoff of the continuum theory. In the following we assume that $\alpha$ is given by the lattice constant of the NWs. 

To incorporate electron-electron interactions, we consider three types of low-energy excitations close to the Fermi surface: (a) $g_4$-type  forward-scattering processes (with momentum transfer $q \sim 0$) coupling fermions only on the same side of the Fermi surface; (b) $g_2$-type forward-scattering processes (with $q \sim 0$) coupling left- and right-moving electrons, however, such that each scattering partner stays on the same side of the Fermi surface, and (c) $g_1$-type backscattering processes (with $q \sim 2 k_F$), where electrons are transferred from one side of the Fermi surface to the other \cite{TG,Senechal}. These scattering processes (a), (b), and (c) [involving two electrons with the same spin], can be incorporated in the kinetic part of the Hamiltonian, whereas the backscattering process (c) involving scattering between electrons with opposite spins should be  considered separately\cite{TG}. 

The kinetic part of  Hamiltonian takes the following form
\begin{align}
H_0= &\sum_{\tau}  \int \frac{dx}{2\pi} \Bigg[
u_{\tau,c} \Big[\frac{\left ( \partial_x \phi_{\tau,c}  \right )^2 } {K_{\tau,c}} + K_{\tau,c} \left ( \partial_x \theta_{\tau,c}  \right )^2\Big] \nn
&\hspace*{1cm}+ u_{\tau,s} \Big[\frac{\left ( \partial_x \phi_{\tau,s}  \right )^2 } {K_{\tau,s}} + K_{\tau,s} \left ( \partial_x \theta_{\tau,s}  \right )^2\Big]\,\Bigg].
\end{align}
where $u_{\tau, c/s}$ and $K_{\tau, c/s}$ are the charge-spin velocity and LL parameters for $\tau$-NW. 

 The simultaneous backscattering of spin up and spin down electrons in $\tau$-NW \cite{TG}, characterized by the coupling strength $g_\tau$, result in the following term in the total Hamiltonian\cite{TG,TG0} 
\begin{align}
H_g =&\sum_\tau g_{\tau} \int dx \left(R_{\tau 1}^\dagger L_{\tau 1} L_{\tau\bar 1}^\dagger R_{\tau\bar1} + \text{ \text{H.c.}}\right)\nn
  =&\sum_\tau 
\frac{  g_{\tau} }{ 2 \, \pi^2 \, \alpha^2 } 
\int dx \cos \left( 2\sqrt{2} \phi_s \right).
\end{align}
To simplify further calculations, we introduce the new bosonic field basis defined as 
\begin{align}
& \phi_{1/2} = \frac{\theta_{1/\bar 1,c}- \phi_{1/\bar 1,s} } {\sqrt 2 }\,,\quad
  \theta_{1/2} = \frac{\phi_{1/\bar 1,c}-\theta_{1/\bar 1,s} } {\sqrt 2 }\,,\nn
& \phi_{3/4} = \frac{  \phi_{1/\bar 1,s}+ \theta_{1/\bar 1,c}} {\sqrt 2 }\,,\quad
  \theta_{3/4} =\frac{\theta_{1/\bar 1,s}+ \phi_{1/\bar 1,c} } {\sqrt 2 }\,.
 \label{nfields}
\end{align}
Expressing the Hamiltonian $H_0$ in this basis we assume that the off-diagonal terms  can be neglected (they are marginally relevant, see below), thus we keep only the diagonal terms, yielding
\begin{align}
&H_0= \sum_{i} u_{i}  \int \frac{dx}{2\pi} 
\Big[\frac{\left ( \partial_x \phi_{i}  \right )^2 } {K_{i}} + K_{i} \left ( \partial_x \theta_{i}  \right )^2\Big]\,,
\label{H0_new}
\end{align}
where $u_i$ and $K_i$ are the new velocity and LL parameters of the NWs. In Appendix \ref{alp},  we derive these LL parameters in terms of original charge ($K_{\tau, c}$) and spin ($K_{\tau,s}$) LL parameters and Fermi velocity $v_{F,\tau}$ of the $\tau$-NW.
Using the relation $u_{\tau,c/s}=v_{F,\tau}/K_{\tau,c/s}$ valid for ideal LLs, we arrive at
\begin{align}
u_{1/2}=&u_{3/4}= \frac{v_{F,1/\bar 1}\sqrt{( 1+K_{1/\bar 1c}^2)( 1+K_{1/\bar 1s}^2)}}{2 K_{1/\bar 1c} K_{1/\bar 1s}}, \label{u}\\ 
K_{1/2}=&K_{3/4}= \frac {K_{1/\bar 1s}}{K_{1/\bar 1c}}\sqrt{(1+K_{1/\bar 1c}^2)/(1+K_{1/\bar 1s}^2})\,. \label{K}
\end{align}
The intrawire and interwire proximity-induced superconducting pairing terms of the Hamiltonian under bosonization reduce to the following form
\begin{align}
&H_s  = \frac{ \Delta_1^{ext}} {\pi\, \alpha}
\int dx \, \cos \left( 2 \,  \phi_1 \right)
+ \frac{ \Delta_{\bar 1 }^{ext}} {\pi\, \alpha}
\int dx \, \cos \left( 2 \,  \phi_2 \right) \nn
& \hspace{0.1 cm}
+    \frac{ \Delta_1^{int}} {\pi\, \alpha}
\int dx \, \cos \left( 2 \, \phi_3 \right) 
+ \frac{ \Delta_{\bar 1}^{int}} {\pi\, \alpha}
\int dx \, \cos \left( 2 \, \phi_4 \right),\label{Hs_new}\\
&H_c   =  \frac{ 2 \Delta_c } {\pi\, \alpha}
\int dx \, \cos \left( \phi_3 + \phi_4\right)\cos \left( \theta_3 -\theta_4\right).
\label{Hc_new}
\end{align}
Notably, in the new basis of bosonic fields, the $\Delta^{ext}_{\tau}$-part commutes with the $\Delta_c $-part and thus they do not compete with each other to form an ordered phase. However, the $\Delta^{int}_{\tau}$-part does not commute with the $\Delta_c $-part and thus they cannot be ordered simultaneously\cite{TG,Oreg_para,JK6}. 
 Finally, the assumption of considering $H_0$ diagonal is justified since the non-diagonal terms are marginal operators in the sense that they are negligible under the RG flow  compared to the cosine terms which flow to their strong coupling regime much faster \cite{Braunecker}.

In the new basis, the Hamiltonian $H_g$ corresponding to the processes of simultaneous backscattering of spin up and spin down electrons in each NW converts to the following form
\begin{align}
&H_g =
\frac{  1 }{ 2 \, \pi^2 \, \alpha^2 } 
\int dx \Big[
 ~g_{1}\cos \{2\left( \phi_1 -\phi_3 \right)\} 
 \nn
 &\hspace*{3.6cm} 
 + g_{\bar 1}\cos \{2 \left(\phi_2 - \phi_4 \right)\}
\Big] \,.
\label{Hg_new}
\end{align}

\section{Treatment of disorder}\label{dis}

In this section, we incorporate the effects of bulk non-magnetic disorder\cite{TG,TG0} in each of the NWs by introducing the term
\begin{align}
H_{dis}= \sum_{\tau} \int dx\ V_\tau(x)\, \rho_\tau(x),
\label{H_dis}
\end{align}
where $V_\tau(x)$ is a random  potential produced by impurities or defects and $\rho_\tau(x)$ is the electron density in the $\tau$-NW. We consider the case of weak uncorrelated disorder following a Gaussian distribution
\begin{align}
\langle V_\tau  (x) \, V_{\tau'}   (x')\rangle _{avg}= D_{\tau} \, \delta(x-x')\, \delta_{\tau\tau'}.
\end{align}
The Gaussian disorder corresponds to the limit of very dense impurities, where  the effect of a single impurity is very weak. The parameter $D_\tau$ measures the strength of the disorder induced by the dense distribution of impurities in the $\tau$-NW. We assume that the disorder in each NW is independent of the other one and the disorder strength $V_{\tau}(x)$ in each NW is much smaller than the Fermi energy such that the disorder affects only the states close to the Fermi points.  In this case, we can focus on Fourier harmonics of the disorder term $V_{q, \tau}$ corresponding to momentum values close to $q\sim 0$ and to $q\sim \pm 2 k_{so,\tau}$, the so-called forward (backward) scattering contributions in each $\tau$-NW. As a result, the Hamiltonian describing disorder takes the following form  
\begin{align}
H_{dis}= & \frac{1}{L}\sum_{\tau,\sigma}\sum_{q\sim 0} V_{q,\tau} \sum_k \psi^\dag_{k+q,\tau\sigma}\psi_{k,\tau\sigma} \nonumber\\
&\hspace{10pt}+
\frac{1}{L}\sum_{\tau,\sigma}\sum_{q\sim \pm 2k_{so,\tau}} V_{q,\tau} \sum_k \psi^\dag_{k+q,\tau\sigma}\psi_{k,\tau\sigma},
\end{align}
which  reduces in the continuum limit to
\begin{align}
H_{dis}= & \sum_{\tau,\sigma}\int dx \, \eta_\tau(x)\, [ R^\dagger_{\tau \sigma}R_{\tau\sigma}+ L^\dagger_{\tau \sigma}L_{\tau\sigma}]\nonumber\\
&\hspace{20pt}+\sum_{\tau,\sigma}\int dx \,[\xi_\tau(x)L^\dagger_{\tau\sigma} R_{\tau\sigma}+  \text{H.c.}],
\label{ldis}
\end{align}
 such that
\begin{align}
\eta_\tau(x)&=\frac{1}{L}\sum_{q\sim 0} V_{q,\tau} \,e^{iqx}\,, \nn
\xi_\tau(x)&=\frac{1}{L}\sum_{q\sim 0} V_{(q-2k_{so,\tau}),\tau} \,e^{iqx}\,.
\end{align}
Here, $\eta_\tau $ and  $\xi_\tau$ ($\xi^*_\tau$) correspond to the $q=0$ and $q=-2 k_{so,\tau}$ ($q=2 k_{so,\tau}$) Fourier components of the random potential $V_\tau(x)$,  respectively. These are essentially independent fields and when averaging over disorder we can use the following relations
\begin{align}
&\langle \eta _\tau  (x) \, \eta_{\tau'}   (x')\rangle _{avg} =D_{\tau} \, \delta(x-x')\, \delta_{\tau\tau'} ,\nonumber \\
&\langle \xi^* _\tau  (x) \, \xi_{\tau'}   (x')\rangle _{avg} = D_{\tau} \, \delta(x-x') \, \delta_{\tau \tau' }, \nonumber \\
&\langle \xi _\tau  (x) \, \eta_{\tau'}   (x')\rangle _{avg} =0,\  \langle \xi _\tau  (x) \, \xi_{\tau'}   (x')\rangle _{avg} =0. \label{avg}
\end{align}
 In terms of the bosonized fields, Eq. (\ref{ldis}) takes the form
\begin{align}
&H_{dis}=  \sum_{\tau}\int dx \,  \frac{\sqrt{2}}{\pi}\big[-\eta_\tau(x) \nabla\phi_{\tau,c} \big]\nonumber\\
&+\sum_{\tau}\int dx \,\big[\, \frac{\xi^*_\tau(x)}{ \pi \alpha} e^{i \sqrt{2}\phi_{\tau,c}} \cos(\sqrt{2}~\phi_{\tau,s})+  \text{H.c.}\,\big].
\end{align}
  We gauge away the forward scattering term by the following transformation:
\begin{equation}
\tilde\phi_{\tau,c}(x)=\phi_{\tau,c}(x)-\frac{ K_{\tau,c}\sqrt{2} }{u_{\tau,c}}\int\limits^x dy  \, \eta(y) \, .
\end{equation}
The only effect of this transformation is to redefine the phase of the backscattering term. The backscattering term leads to pinning of the fields, which corresponds to localization in one-dimension systems. Moreover, in order to deal with the disorder averaging, we use the replica method \cite{TG,TG0}. We introduce $N$ copies of the fields $(\phi_{i},\theta_{i})\to 
( \phi^n_{i},\theta^n_{ i })$
with $n \in [1,N ]$, average over the Gaussian disorder, and finally take the limit $N\to 0$. At the end, we obtain a Gaussian action in the replica space which we use to derive the RG equations. The replica term in the action, obtained with help of Eq. (\ref{avg}), is given by
\begin{align}
\label{disorderham}
& S_{dis,\tau} =\frac{-D_\tau } { 2  \pi^2 \alpha^2 } \Bigg[\sum_{m,n}  \int dx \,dt \, dt' 
e^{i \sqrt{2} \phi^m_c(x,t)} e^{-i \sqrt{2} \phi^n_c(x,t')}
\nn &\times\cos \left( \sqrt 2 \,  \phi^m_{\tau , s } (x,t )  \right )  \cos \left(  \sqrt 2 \,  \phi^n_{ \tau, s  } (x,t') \right )\, +\text{H.c.}
\Bigg],
 \end{align}
where $m$ and $n$ are replica indices,  while $t$ and $t'$ are imaginary time coordinates.  We rewrite the foregoing action for each NW in terms of the new fields given by Eq. (\ref{nfields}) as
 \begin{align}
& S_{dis,1/\bar 1} =  -\frac{D_{1/\bar 1} } { 2\, \pi^2 \alpha^2 } \Bigg[\sum_{m,n} \int dx \,dt \, dt'
e^{i \big\{ \theta^m_{ 1/2 } (x,t ) + \theta^m_{ 3/4 }(x,t)\big\}} \nn
& \times e^{-i\big\{  \theta^n_{ 1/2 } (x,t' ) + \theta^n_{ 3/4 }(x,t')\big\}}
 \cos \big\{   \phi^m_{ 1/2 } (x,t ) - \phi^m_{ 3/4 } (x,t) \big\} \nn 
& \hspace{10pt}\times \cos \big\{  \phi^n_{ 1/2 } (x,t') - \phi^n_{ 3/4 } (x,t' ) \big\} +\text{H.c.}\Bigg].\label{Hdis_new}
\end{align} 
Below, we calculate the RG flow equations in first order in $D_\tau$, therefore the perturbative expansion will be carried out without the replica indices \cite{TG,TG0}. 

\section{RG equations for effective Hamiltonian}\label{rg}

In the following section, we investigate the RG flow equations for different parameters in the system. Collecting all terms described above, we define an effective Hamiltonian for a double-NW setup with electron-electron interaction as
\begin{align}
H_{eff}= H_0+H_s+H_c+H_g,
\label{Heff}
\end{align}
where $H_0$, $H_s$, $H_c$, and $H_g$ are given by Eqs. (\ref{H0_new}), (\ref{Hs_new}), (\ref{Hc_new}), and (\ref{Hg_new}), respectively. 
To the action obtained from $H_{eff}$ we add the non-local contribution coming from the disorder averaged part given in Eq. (\ref{Hdis_new}).
The proximity induced pairing and disorder terms  in the Hamiltonian are competing with each other since they do not commute, therefore we perform a standard RG analysis \cite{TG} to find out which terms are dominant as a function of the system parameters. In Ref. \onlinecite{Suhas2}, it has been shown that in the presence of the interactions, velocities flow to the equal limit, thus we do not incorporate the renormalization of velocities while deriving the RG equations for the couplings. This amounts to assuming that $u_{i}\equiv u$.
In what follows, we use the dimensionless coupling constants defined as $\tilde \Delta_{\tau /c}=\Delta_{\tau/c}\, \alpha/u$, $\tilde D_\tau= \alpha \, D_\tau/(2\pi u^2)$, and $y_\tau=g_\tau/(2\pi u)$. 

From the RG flow equations, we can deduce the conditions under which the interwire (crossed Andreev) pairing gap dominates over the intrawire one and over the disorder. In the RG interpretation, this means to find a parameter regime for which $\tilde\Delta_c $ becomes of order one, while $\tilde\Delta^{int/ext}_\tau< \tilde\Delta_c$, and $\tilde D_\tau$, $y_\tau \ll 1$.  When these conditions are satisfied, the spectrum is gapped [see Eq.~(\ref{topocrit})] and the system is in a topological phase supporting KMBSs. We note that this bosonic system can be adiabatically connected to the non-interacting system as was done in Ref. [\onlinecite{Suhas}] since the relevant LL parameter $K_3$ flows to the effectively non-interacting value, thus allowing a refermionization of the action to a quadratic Hamiltonian but with all gaps renormalized by interactions.
 \begin{figure}[t]
        \centering
                \includegraphics[width=0.4 \textwidth]{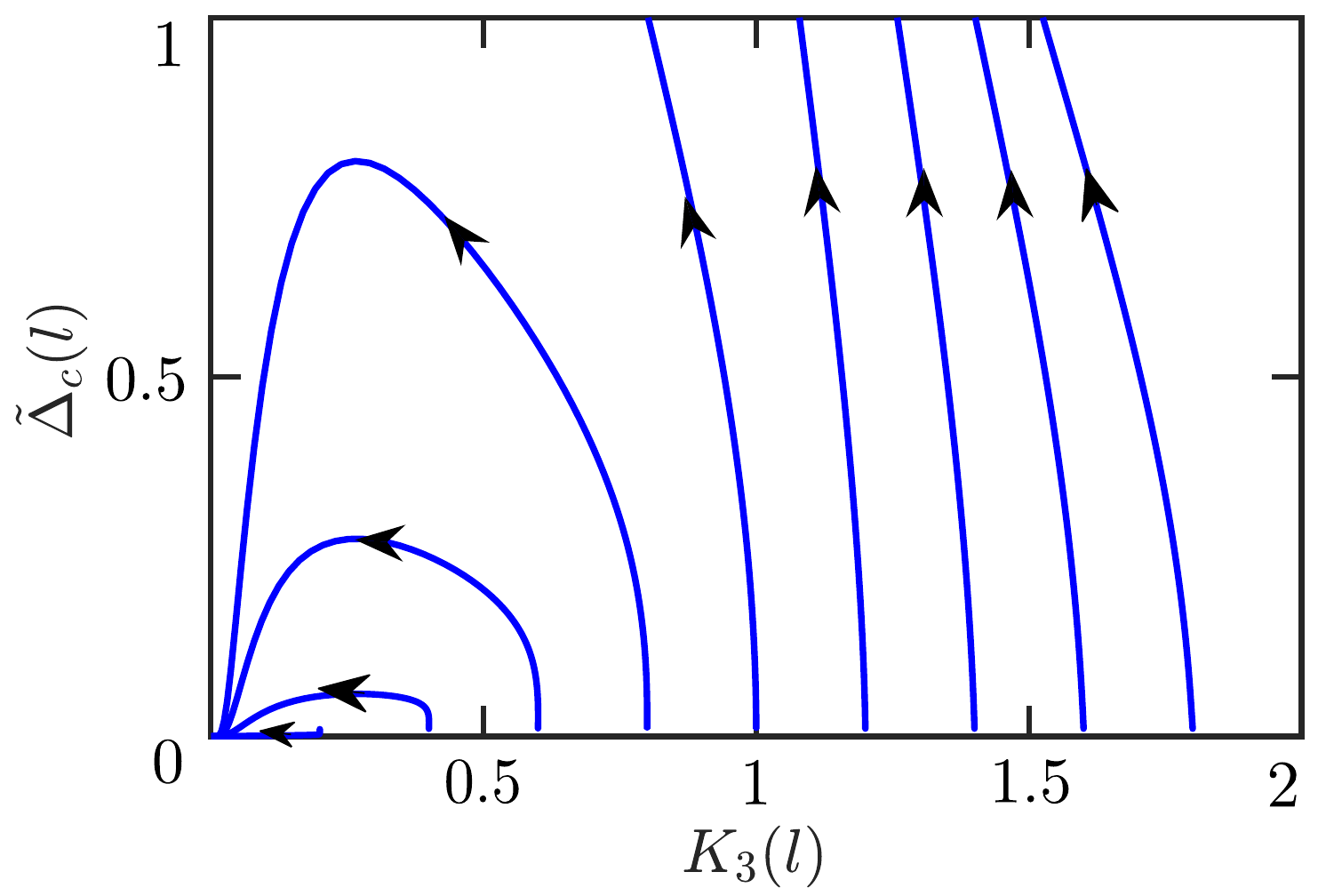} 
                 \caption{\label{rgplot2} The RG flow of crossed Andreev pairing $\tilde \Delta_c$ as a function of LL parameter $K_3$, see Eq. (\ref{rg1}). The initial conditions are changed from $K_1(0)=K_3(0)=0.2$ to $1.8$ (the most right flow line) with a step 0.2. As expected, $\tilde\Delta_c=0$ is the stable fixed point, which is reached for $K_3(0)<1$. If $K_3(0)>1$, $\tilde\Delta_c$ reaches the strong coupling limit, $\tilde\Delta_c=1$, after which we stop the RG flow. The rest initial conditions are fixed to  $\tilde   \Delta_{ \tau }^{ext}(0)  = \tilde \Delta_{ \tau }^{int}  (0) = \tilde \Delta_c(0)=0.01$ and $\tilde D(0)=y(0)=0.001 $.  }
\end{figure}
To derive the RG equations for the coupling constants and LL parameters present in the effective Hamiltonian given by Eq. (\ref{Heff}), we use the operator product expansion (OPE) expressions \cite{{Cardy,Schoeller,Senechal,Tsuchiizu}} listed in Appendix \ref{aope}. As a result, we arrive at [see Appendix \ref{arg} for more details] 
\begin{align}
\label{rg1}
& \frac{d \tilde \Delta_{ 1 }^{ext}  }{ dl } = 
\Big [ 2 -K_1  \Big ] \tilde \Delta_{1}^{ext},\quad
 \frac{d \tilde \Delta_{ \bar 1 }^{ext}  }{ dl } = 
\Big [ 2 -K_2  \Big ] \tilde \Delta_{\bar 1}^{ext},\nn
& \frac{d \tilde \Delta_{ 1 }^{ int }  }{ dl } = 
\Big [ 2 -K_3  \Big ] \tilde \Delta_{1}^{int } ,\quad
 \frac{d \tilde \Delta_{ \bar 1 }^{int}  }{ dl } = 
\Big [ 2 -K_4  \Big ] \tilde \Delta_{\bar 1}^{ int },\nn
& \frac{d \tilde \Delta_{  c }   }{ dl } = 
\Big [ 2 -\frac{1}{4}  \left ( K_3 + K_4 +\frac{1} {K_3}   + \frac{1} {K_4}  \right ) \Big ] \tilde \Delta_{ c },\nn
& \frac {dK_{1}}{dl}
= 
- \big[ \left ( \tilde \Delta_{ 1 }^{ext} \right)^2  +   y_1^2\big]    K_{ 1  }^2 +\frac{\tilde D_1 ( 1- K_{ 1 }^2 )} { 2} ,\nn
& \frac {dK_{ 2 }}{dl}
= 
- \big[  \left ( \tilde \Delta_{\bar  1 }^{ext} \right)^2 +   y_{\bar  1 }^2 \big]  K_{ 2  }^2  +\frac{\tilde D_{\bar 1} ( 1- K_{ 2 }^2 )} { 2},\nn
& \frac {dK_{ 3  } }{dl}
= -  \big[ \left ( \tilde \Delta_{ 1 }^{int} \right)^2 + y_1^2   \big]  K_{ 3 }^2 
 + \frac{ (\tilde\Delta_c^2+\tilde D_1) \, ( 1- K_{ 3 }^2 )} { 2},\nn
 &
 \frac {dK_{ 4  } }{dl}
= -  \big[ \left ( \tilde \Delta_{ \bar 1 }^{int} \right)^2+ y_{\bar  1 }^2   \big]  K_{ 4 }^2  + \frac{ (\tilde \Delta_c^2+\tilde D_{\bar 1}) \, ( 1-  K_{ 4 }^2 )} { 2},\nn
& \frac{ d \tilde D_1 }{ dl } = \Big [  3 -  \frac{1}{2}  \left ( K_1 + K_3 +\frac{1} {K_1}   + \frac{1} {K_3}  \right ) 
-y_1 \Big ] \tilde D_1 , \nn
& \frac{ d \tilde D_{\bar 1} }{ dl } = \Big [  3 -  \frac{1}{2}  \left ( K_2 + K_4 +\frac{1} {K_2}   + \frac{1} {K_4}  \right ) 
-y_{\bar 1} \Big ] \tilde D_{\bar 1}, \nn
& \frac{dy_1}{dl} = (2- K_1-K_3 )\, y_1 - \tilde D_1  , \nn
& \frac{dy_{\bar 1}}{dl} = (2- K_2-K_4 )\, y_{\bar 1} - \tilde D_{\bar  1 } , 
\end{align}
where $l=\ln(\alpha/\alpha_0)$ is the dimensionless RG flow parameter, $\alpha_0$  is the lattice constant of the NWs, and $\alpha$ is the rescaled lattice constant that grows under the various perturbations. 

\begin{figure}[t]
        \centering
         {\includegraphics[width=0.4 \textwidth]{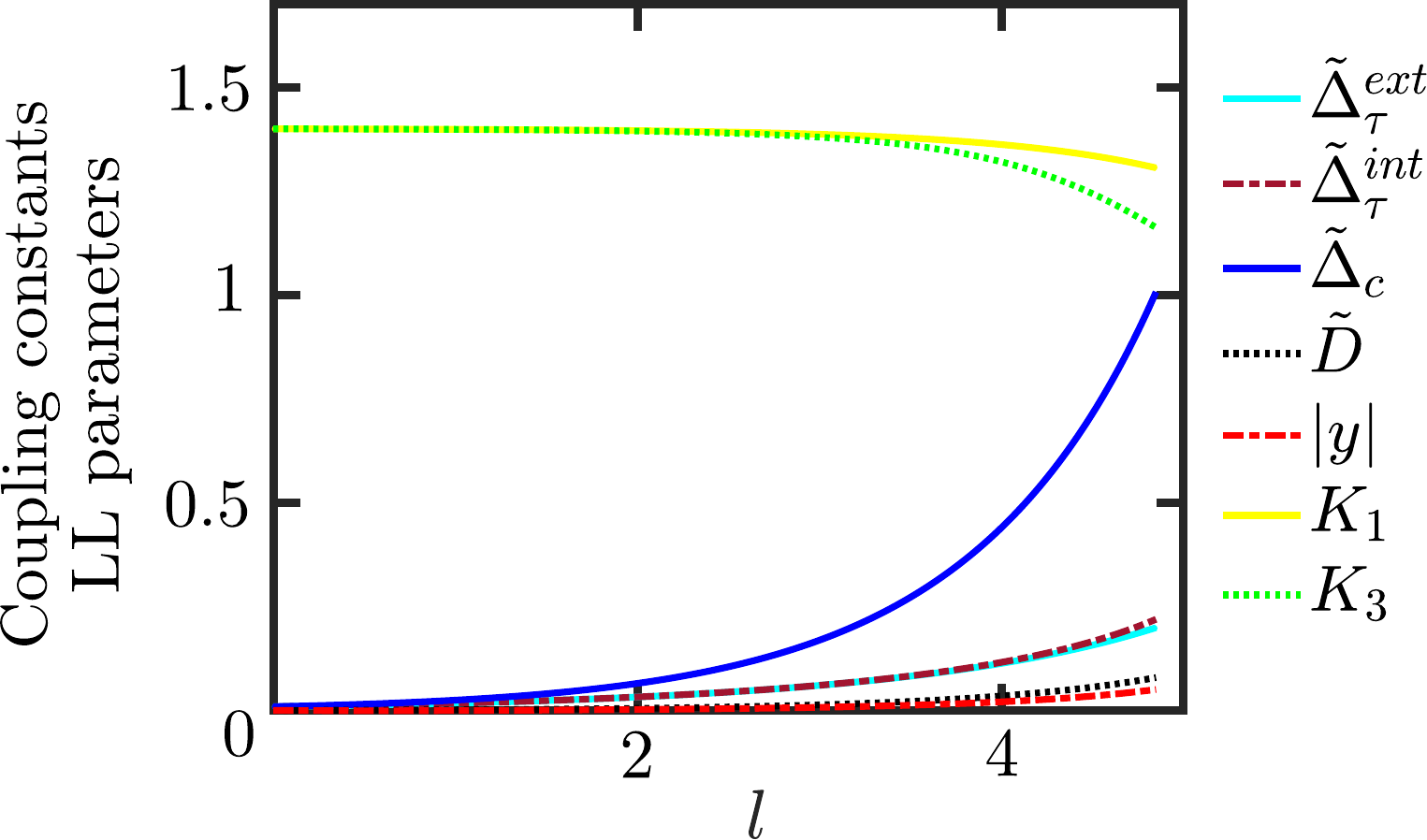}}
        \caption{\label{rgplot1} The RG flow of various dimensionless coupling constants and LL constants for initial conditions $\tilde   \Delta_{ \tau }^{ext}(0)  = \tilde \Delta_{ \tau }^{int}  (0)= \tilde \Delta_c(0)=0.01$, $\tilde D(0)=y(0)=0.001$, and $K_1(0)=K_3(0)= 1.4 $, see Eq. (\ref{rg1}). The crossed Andreev pairing $\tilde \Delta_c$ (blue solid) grows much faster than the intrawire  pairing $\tilde\Delta_\tau^{int}$ (brown dashed) or the disorder term strength $\tilde{D}$ (black dotted) under the flow parameter $l$,  enabling the topological phase for given initial conditions.  We note that  $y$ (red dashed) flows to a negative value for $K_i(0)>1$, so we plot the absolute value  $|y|$ \cite{TG}. }
\end{figure}

\begin{figure}[t]
        \centering
                \includegraphics[width=0.4\textwidth]{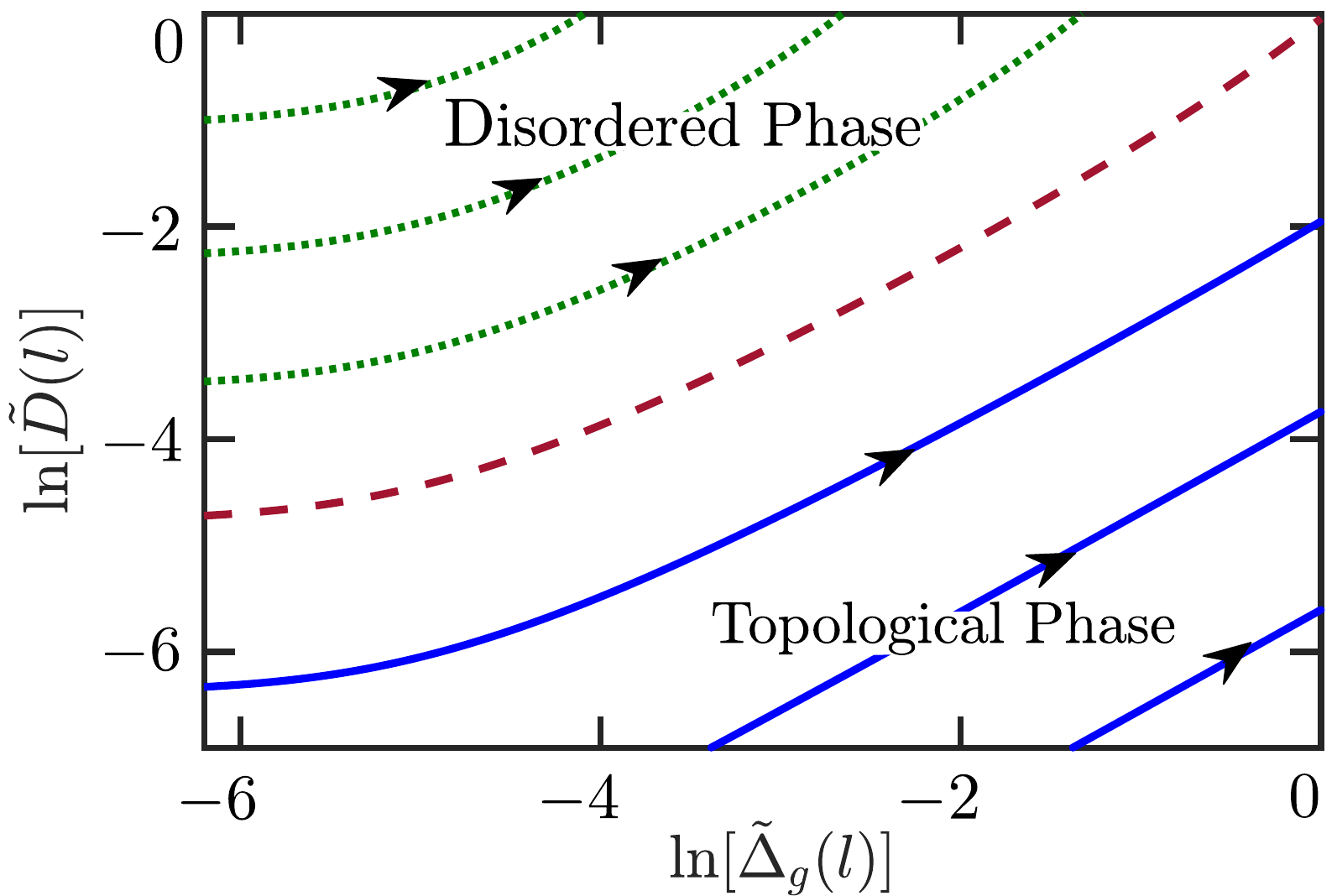} 
        \caption{\label{rgplot3} The flow phase diagram consisting of topological and disorder phases obtained numerically from  Eq. (\ref{rg1}). Both disorder strength  $\tilde D$ and  the topological gap  $\tilde \Delta_g=\sqrt{\tilde \Delta_c^2-\tilde \Delta_1^{int} \tilde \Delta_{\bar 1}^{int}}$ are increasing under the flow.  By comparing $\tilde \Delta_g$ and   $\tilde D$  on a log-log scale with different initial conditions for $\tilde \Delta_g (0)$ and $\tilde D (0)$, we define the topological (disordered) phase as one in which  $\tilde \Delta_g$ ($\tilde D$) reaches the strong coupling regime first. The red dashed curve separates the two phases and is defined by a condition that both $\tilde \Delta_g$ and $\tilde D$ reach the strong coupling limit simultaneously. In the regime of weak (strong) disorder and initially large (small) topological gap, the system is in the topological (disordered) phase indicated by blue solid (green dotted) flow lines.
The other initial parameters are fixed to $K_1(0)=K_3(0)=1.4$, $\tilde   \Delta_{ \tau }^{ext}(0)  = \tilde \Delta_{ \tau }^{int}  (0) = 0.01 $, and $y(0)=0.001$.  
 }
\end{figure}

To reduce the number of parameters, we assume that the electron-electron interactions are similar in both NWs so we set $K_1=K_2$ and $K_3=K_4$ such that there are only seven independent parameters in Eq. (\ref{rg1}).  Even with these assumptions, the system of coupled RG equations stays involved and below we comment on limiting cases. 

For $\tilde{\Delta}_\tau^{ext}$, $\tilde{\Delta}_\tau^{int}$, $\tilde \Delta_c$ to be relevant in the RG sense (terms grow with $l$), we should have $K_1<2$, $K_3<2$, and $K_3+K_3^{-1}<4 \Rightarrow (2-\sqrt{3})<K_3< (2+\sqrt{3})$, respectively.  Thus, for $(2-\sqrt{3})<K_3<2$, both $\tilde \Delta_c$  and $\tilde{\Delta}^{int}_\tau$ are relevant. 
For the disorder coupling constant $\tilde D$ to be a relevant parameter, the condition $K_1+K_3+K_1^{-1}+K_3^{-1}+y_\tau<6$ should be satisfied. Motivated by the initial condition of LL parameters (see Appendix \ref{alp}), $K_1(0)=K_3(0)$, for our estimate we use $K_1=K_3$ and $y_\tau \rightarrow 0$. Thus, disorder is a relevant parameter in the range $(3-\sqrt{5})/2< K_{1,3}<(3+\sqrt{5})/2$  of LL parameters. For the backscattering coupling constant $y$ to be a relevant parameter in the absence of disorder ($\tilde D \rightarrow 0$), one needs to work in the regime $K_1+K_3 < 2$.  Generally, in the most interesting topological regime, both disorder and backscattering terms can be relevant. Thus, we consider an interplay between these and the superconducting terms taking into account also the initial values of the coupling constants and LL parameters, which determine the RG flow.
 \begin{figure}[t]
        \centering
                \includegraphics[width=0.36 \textwidth]{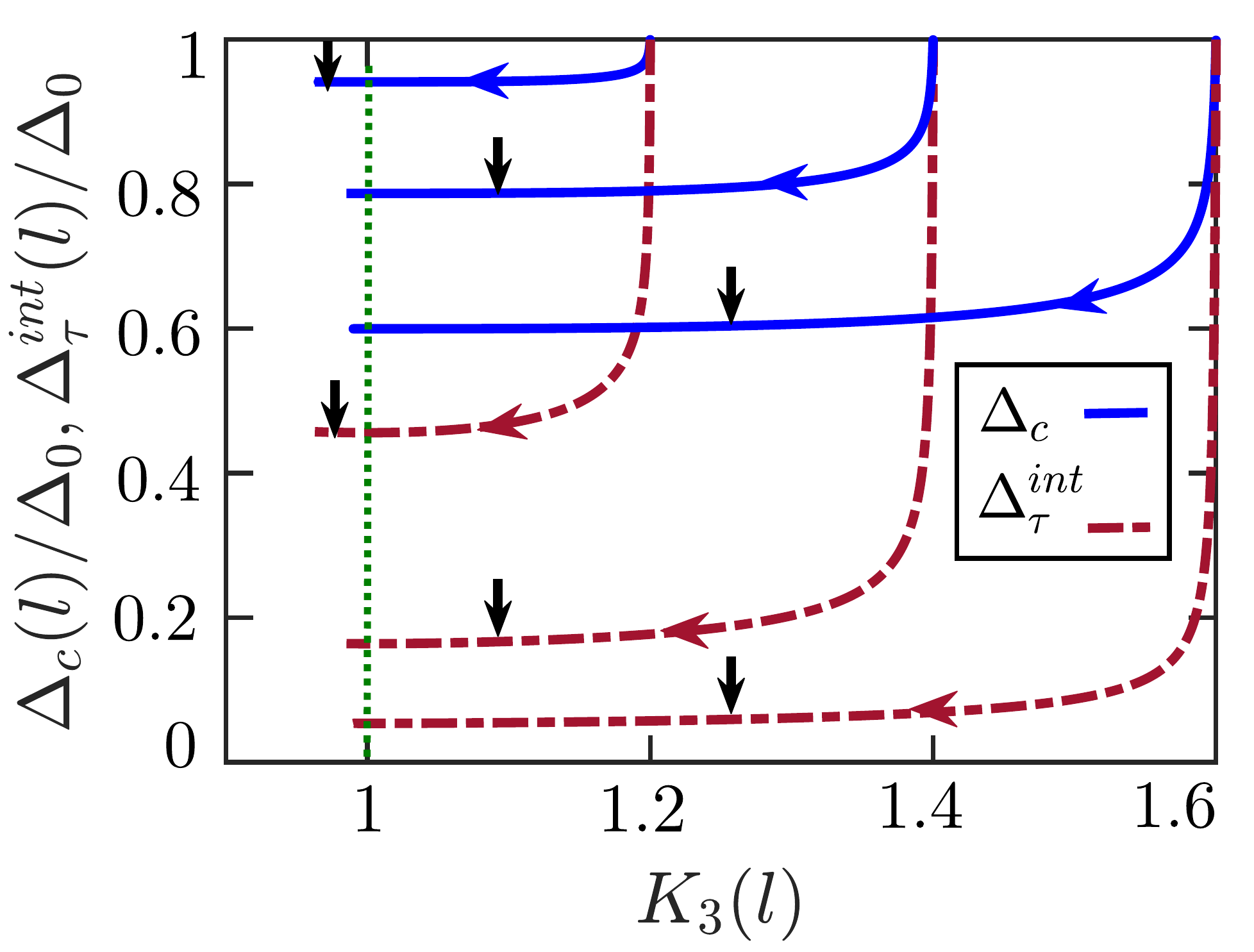} 
        \caption{\label{rgplot0} The RG flows for the physical (dimensionful) pairing terms $\Delta_c/\Delta_0$ (blue solid curves) and $\Delta_\tau^{int}/\Delta_0$ (red dashed curves) as a function of $K_3$ obtained from Eq. (\ref{rg1}) with three different initial conditions: $K_1(0)=K_3(0)=1.2$, $1.4$, and $1.6$ (from left to right flow lines), which corresponds to $K_c=0.73$, $0.59$, and $0.49$, respectively for $K_s=1$, see Eq. (\ref{K}). The vertical black arrow corresponds to the point where $ \tilde \Delta_c$ =1 is reached. The vertical green dotted line at  $K_3=1$ corresponds to the non-interacting limit, where the bosonic system can be refermionized to a non-interacting system with renormalized pairing gaps~\cite{Suhas}.  For $K_3>1$, the value of interwire pairing gap is always greater than the respective intrawire pairing gap, hence, the system hosts KMBSs at each end of the NWs. 
 The other initial conditions are fixed to $\Delta_0=  \Delta_{ \tau }^{ext}(0)  = \Delta_{ \tau }^{int}  (0) =  \Delta_c(0)=0.01\, u/a_0$, $\tilde D(0)=y(0)=0.001 $. }
\end{figure}

 In what follows we will focus on $\tilde \Delta_c(l)$ and find regimes in which it dominates over other coupling constants. First, we solve the system of coupled RG equations to find the parametric dependence of $\tilde \Delta_c(l)$ on $K_3(l)$ for various initial conditions of LL parameters $K_1(0)=K_3(0)$, see Fig. \ref{rgplot2}.
 We find that $\tilde \Delta_c=0$ is the stable fixed point. Moreover, for the initial condition $K_1(0)=K_3(0)< 1$, the crossed Andreev pairing amplitude $\tilde \Delta_c$ flows to zero before reaching the strong coupling limit. In contrast to that, for $K_1(0)=K_3(0)> 1$, $\tilde \Delta_c$ reaches the strong coupling limit as long as the initial values of $\tilde D$ is smaller than $\tilde \Delta_c$. We note here that for $K_{1,3}>1$, the backscattering coupling constant $y$ is an irrelevant parameter and flows to negative values. We remind that the flow is stopped as soon as $\tilde \Delta_c=1$. The RG flow of all other coupling constants and LL parameters in the regime of interest $K_1(0)=K_3(0)> 1$ as a function of flow parameter $l$ is shown in Fig. \ref{rgplot1}. Indeed, the crossed Andreev term grows most rapidly and reaches the strong coupling limit first for some range of parameters and thus drives the system into the topological phase (see below). We note that in spite of the fact that the RG flow equations for $K_1$ and $K_3$ are different, they stay almost the same during the flow, which justifies the estimates of scaling dimensions done above. The LL parameters $K_{1,3}$ can be mapped back to more standard $K_{\tau c}$ and $K_{ \tau s}$ by using Eq. (\ref{K}). 
To enter the topological phase, we need $K_{1,3}(0)>1$ which implies $K_{\tau c}<K_{ \tau s}$. If the interactions are such that the spin rotation symmetry is preserved (broken), $K_{\tau s}=1$ ($K_{\tau s}>1$). In any case, $K_{\tau c}$ is always smaller than one, which corresponds to repulsive interactions.  To analyze the stability of the obtained topological phase, we explore different initial values of coupling constants and obtain the phase diagram.
  \begin{figure}[t]
        \centering
                \includegraphics[width=0.36 \textwidth]{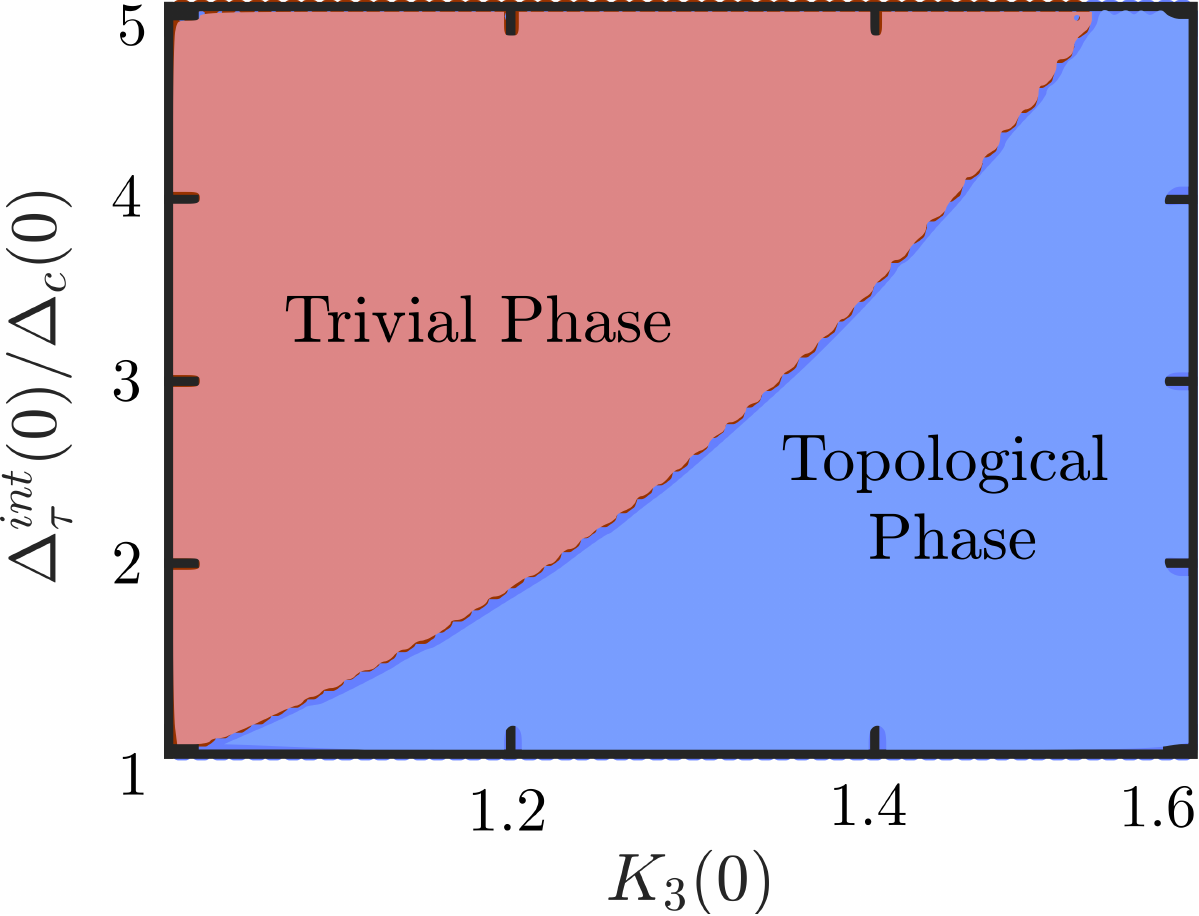}
                \caption{\label{phase} Phase diagram determined numerically from the RG flow for different initial values of $\Delta_\tau^{int}(0)/\Delta_c(0)$  and $K_3(0)$. At the end of the RG flow (when one of the coupling constants reaches one), the pairing amplitudes satisfy $\Delta_c (l)> \Delta_\tau^{int}(l)$ in the topological phase (blue area).  In the trivial phase (red area), $\Delta_c (l)< \Delta_\tau^{int}(l)$.
Remaining initial conditions are $\tilde D(0)=y(0)=0.001 $ and $K_1(0)=K_3(0)$.}
\end{figure} 
First, we focus on effects of disorder on the topological phase. The RG flow of the coupling constants and the LL parameters shown in Fig. \ref{rgplot1} indicate that for the  initial values $K_{1,3}(0)>1$, though disorder grows fast under RG, as long as the starting value of $\tilde D (0)$ is small enough, the crossed Andreev  pairing amplitude $\tilde \Delta_c$ reaches the strong coupling limit first before the disorder can grow to an appreciable value. However, it is expected that if disorder is strong initially, it will win over the superconducting gaps and drive the system into the disordered phase without MBSs.  We solve numerically the coupled RG equations [see Eq. (\ref{rg1})] for different initial conditions of disorder strength $\tilde D$ and of the topological gap $\tilde\Delta_g=\sqrt{\tilde \Delta_c^2-\tilde \Delta_1^{int} \tilde \Delta_{\bar 1}^{int}}$ for $K_{1,3}>1$, see Fig. \ref{rgplot3}.   Here, we assume that the system is already in the topological phase with $\tilde \Delta_c^2(0)>\tilde \Delta_1^{int}(0) \tilde \Delta_{\bar 1}^{int}(0)$.  Under the RG, both $\tilde\Delta_g$ and $\tilde D$ flow to larger values.  If $\tilde D$ ($\tilde\Delta_g$) reaches the strong coupling limit first, the system is in the disordered (topological) phase. The critical line indicating the quantum phase transition separating the two phases is defined by the condition that both couplings $\tilde D$ and $\tilde\Delta_g$ reach the strong coupling limit simultaneously. 
We conclude that in order to reach the topological phase, one should have $K_{1,3}(0)>1$ and small initial values of $\tilde D (0)$ compared to the topological gap $\tilde\Delta_g (0)$.

We have found that in the topological phase the dimensionless coupling constants $\tilde \Delta_c$ and $\tilde \Delta_{\tau}$ are always increasing under the RG flow, see Figs. \ref{rgplot1} and \ref{rgplot3}. However, one should keep in mind that for realistic systems only the physical values of the superconducting pairing $\Delta_{\tau/c}$ are of relevance.  To extract these physical gaps, we use the relation $\Delta_{\tau/c}= \tilde \Delta_{\tau/c}u/\alpha$.  By plotting numerically the flow lines of $\Delta_{\tau/c} (l)$  versus the LL parameters $K_3(l)$ (see Fig. \ref{rgplot0}), we see that  electron-electron interactions suppress the superconducting gaps, which is a generic behavior as discussed in Ref.  [\onlinecite{Suhas}]. Importantly, the intrawire superconducting pairing is suppressed stronger than the interwire (crossed Andreev) superconducting pairing. This reflects the physical expectation that interactions suppress the tunneling of two electrons  in the same NW stronger than when the two electrons from the Cooper pair separate and tunnel each into a different NW. Following the reasoning of Ref. [\onlinecite{Suhas}], we continue the flow until the special point $K_{1,3}=1$ (effectively non-interacting limit) is reached. At this point we can refermionize our bosonic system to an effectively non-interacting system whose Hamiltonian is purely quadratic in the fermionic operators, and solve for the spectrum straightforwardly with renormalized pairing amplitudes given at $\Delta_c(K_3=1)$ and $\Delta_\tau(K_3=1)$. For these parameters, we find that the topological criterion Eq. (\ref{topocrit}), $\Delta_c^2>\Delta_1^{int}  \Delta_{\bar 1}^{int}$, is satisfied. Hence, the system is in the topological phase and hosts one KMBS at each end of the double-NW system. 
We also explored the stability of the obtained topological phase towards initial conditions.  In Fig. \ref{phase}, we plot the phase diagram for different initial values of pairing amplitudes and LL parameters. Importantly, even if  the system is initially in the trivial phase with $\Delta_\tau^{int}(0)>\Delta_c(0)$, under the RG flow, electron-electron interactions drive the system into   the topological phase for which  $\Delta_\tau^{int}<\Delta_c$. However, as the ratio between the initial values of  superconducting pairings $\Delta_\tau^{int}(0)/\Delta_c(0)$ increases, we need increasingly stronger repulsive electron-electron interactions in the NWs to reach the topological phase.

\section{RG treatment of the tunneling Hamiltonian in the source term approach}\label{rgt}

In the previous section we worked with the effective Hamiltonian. In particular,
we included  intra- and interwire superconducting pairing terms in the Hamiltonian as model parameters [see Eqs. (\ref{HSC1}) and (\ref{HSC2})]. Afterwards,  we computed the RG flow by using Eq. (\ref{rg1}).  In a more microscopic approach, one should begin with the tunneling Hamiltonian between the superconductor and NWs, and derive the intrawire (direct) and interwire (crossed Andreev) pairing amplitudes explicitly.
Such an approach has been developed, for example, in Ref. [\onlinecite{Reeg_a}] for the same double-NW setup without electron-electron interactions or disorder. 
Here, we  derive the RG flow equations
for the superconducting pairing amplitudes in the presence of electron-electron interactions with the tunneling term being taken into account as a source term in the RG equations. In doing so, we follow the work of Virtanen and Recher~\cite{Recher} who introduced such a source term formalism to describe proximity-induced superconductivity
in strongly interacting edges of topological insulators. 
In this section, we do not consider the disorder and backscattering terms explicitly as was done in Sec. \ref{rg}.  
However, we have checked disorder effects numerically as discussed below.

 We model the coupling between the three-dimensional bulk $s$-wave superconductor (SC) and the NWs by the following tunneling Hamiltonian,
\begin{align}
&H_T
 =  \sum \limits_{\tau}  \int dx \, d{\bf r} \, 
  \Big\{ \Big [
t'_{ext,\tau} (x, {\bf r})\, e^{ -i \, k_{F\tau} \, x} \, R^\dagger _{\tau 1}(x)\nn
 &~~~~+ t'_{int,\tau} (x, {\bf r}) \,  L^\dagger _{\tau 1}(x)
\Big ] \, \Psi_{\uparrow} ({\bf r} ) +  
\Big [
t'_{int,\tau} (x, {\bf r}) \, R^\dagger _{\tau \bar 1}(x)\nn 
&~~~~+ t'_{ext,\tau} (x, {\bf r})e^{  i \, k_{F\tau} \, x} \,  L^\dagger _{\tau \bar 1 }(x)
\Big ]\, \Psi_{\downarrow} ({\bf r} ) + \text{H.c.} \Big \} , 
\label{HT}
\end{align}
where the operator $\Psi_{\sigma} ({\bf r} )$ is an annihilation operator acting on electrons with spin $\sigma$ located at point $\bf r$ of the SC.
The SC is characterized by the anomalous Green function
\begin{align}
F({\bf r},{\bf r}',\omega)=\int  &  \frac{ d{\bf k}}{(2\pi)^3}\, e^{i( {\bf r}- {\bf r}')\cdot{\bf k}} \frac{\Delta}{\omega^2+E_k^2+\Delta^2},
\end{align} 
where  $\Delta $ is the superconducting pairing amplitude. 
The energy  dispersion of the SC in the normal phase is given by $E_k =  \hbar^2 (k^2- k_{F,sc}^2)/2 m_e$, where $m_e$ and $k_{F,sc}$ are the mass of the electron and the Fermi wavevector of the SC, respectively. We again separate  the tunneling amplitudes into two parts, $t'_{ext}$ and $t'_{int}$, that act at momenta close to $\pm k_{F\tau}$  and zero, respectively, which is important for the correct treatment of $\Delta^{int/ext}_\tau $.  Hence, $t'_{int}$ results in the source terms for generating the intrawire superconducting pairing for the interior branches, $\Delta_\tau^{int}$, and the interwire (crossed Andreev) superconducting pairing, $\Delta_c$, while $t'_{ext}$ results in the source term for generating the intrawire  superconducting pairing for the exterior branches, $\Delta_\tau^{ext}$. Note that for simplicity we set the tunneling strengths equal for both  NWs and  assume the point-like tunneling
\begin{align}
  t'_{int/ext,\tau} (x,{\bf r})=  { t}_{int /ext }    \, \delta( r_x - x)\,\delta(r_y-d_\tau ) \,\delta( r_z ) , \label{tapprox1}
\end{align}
where without loss of generality we assume that $d_1=0$ and $d_{\bar 1}=d$, with $d$ being the distance between two NWs aligned in the $x$ direction and placed in the $xy$-plane, see Fig. \ref{fig00}. 

Similarly to the previous section, before deriving the RG flow equations, we should introduce dimensionless parameters also for the tunneling terms, which will allow us later to define the strong coupling regime.
From dimensional analysis of Eqs. (\ref{HT}) and (\ref{tapprox1}) and noting that $\Psi_{\uparrow/\downarrow}$  is normalized to the volume of the bulk SC,
we see that ${ t}_{int /ext }$ depends on (volume)$^{1/2}$.
We also recall that the system is translationally invariant along the $x$ direction, apart from boundary effects which, however, we ignore for the present RG analysis by assuming that $L$, the length of the NWs in $x$-direction, is much longer than any other length scales. 
Furthermore,  the tunnel contributions of Cooper pairs from the SC to the NWs (responsible for the proximity gaps in the NWs) in the transverse $y$- and $z$-directions can only come from
distances up to the coherence length $\xi= \hbar v_{F,sc}/{\Delta}$ within the SC, where $\Delta$ ($v_{F,sc}$) is the gap (Fermi velocity) of the bulk SC.  In addition, obviously, the proximity-induced superconducting gaps should not depend on the size of the system in $y/z$ direction as long as it  exceeds $\xi$.
Thus, the natural length scales for dividing out the volume dependence of ${ t}_{int /ext }$ is given by $\sqrt{\xi^2\,L}$. As a result, applying dimensional analysis and using Eq.~(\ref{chiralfield}), we confirm that $\frac{t_{int/ext} \sqrt{\alpha}}{(\xi^2\, L)^{1/2}}$ has the dimension of energy.
Thus, we define the dimensionless coupling constants as 
\begin{align}
\tilde{ t}_{int/ext}= t_{int/ext}\sqrt{\frac{\alpha}{\xi^2\,L}} \times \frac{\alpha}{u},
\label{dimensionless_tun}
\end{align}
where $u$ is the Fermi velocity in the NWs. 
Taking again the cut-off  $\alpha$ as the lattice constant of the NWs, we see that $\tilde{ t}_{int/ext}\propto \sqrt{1/N}$, where $N=L/\alpha$ is the number of lattice sites of the NWs. Hence, $\tilde{ t}_{int/ext}$ decreases with increasing $L$ (or $N$), but this decrease is  compensated by the increase of number of states (with increasing $L$) in the superconductor that contribute to the formation of the proximity gaps \cite{Reeg_a}. As a result, as expected, the proximity induced gaps are independent of the length of the system.
For simplicity, we assumed as before that the Fermi velocities $u_i=u$ are the same in both NWs as well as the LL parameters $K_1=K_2$ and $K_3 = K_4$,  and, in addition, we consider only the lowest order terms in $\tilde t_{int/ext}$.
Under these assumptions,  we derive the following set of coupled RG equations (see Appendix \ref{argt}  for more details):
\begin{align}
\label{rg4}
& \frac{d \tilde t_{  int }   }{ dl } = 
\Big [ 2 -\frac{1}{ 4 }  \left ( K_3 +  \frac{1} {K_3}    \right ) \Big ] \tilde t_{  int } \,,
\nn &
\frac{d \tilde t_{ ext }   }{ dl } = 
\Big [ 2 -\frac{1}{ 4 }  \left ( K_1 +  \frac{1} {K_1 }    \right ) \Big ] \tilde t_{  ext }\,, \nn
& \frac{d \tilde \Delta_{ \tau }^{ext}  }{ dl } = 
\Big [ 2 -K_1  \Big ] \tilde \Delta_{\tau}^{ext} +S \,\tilde t^2_{ext}\,,\nn&
 \frac{d \tilde \Delta_{ \tau }^{ int }  }{ dl } = 
\Big [ 2 -K_3  \Big ] \tilde \Delta_{\tau}^{int } +S\, \tilde t^2_{int}\,,\nn
& \frac{d \tilde \Delta_{  c }   }{ dl } = 
\Big [ 2 -\frac{1}{2}  \left ( K_3 +  \frac{1} {K_3}    \right ) \Big ] \tilde \Delta_{ c } + S_c \,\tilde t^2_{int}\,,\nn
& \frac {dK_{1}}{dl}
= 
-  \left ( \tilde \Delta_{ \tau }^{ext} \right)^2      K_{ 1  }^2    \,,\nn &
  \frac {dK_{ 3  } }{dl}
= -   \left ( \tilde \Delta_{ \tau }^{int} \right)^2    K_{ 3 }^2 
 + \frac{\tilde \Delta_c^2 \, ( 1- K_{ 3 }^2 )} { 2} \,.
 \end{align}
The source term  $S \,\tilde t_{int/ext}^2$ governs the intrawire direct pairing processes, while the source term $S_c \,\tilde t_{int}^2$  governs the interwire crossed Andreev pairing processes. Here, $S$ and $S_c$ are given by (see App. \ref{argt})
 \begin{align}
S= &\frac{m_e \,v_{F,sc}^2 \,  L}{ 2\, \pi\,\Delta \, \alpha}  
 K_0\left( \frac{ \alpha \Delta   } {\hbar\, u} \right),
 \label{SS'1}\\
   S_c=&\frac{m_e\, v_{F,sc}^2\,L\,  |\sin(k_{F,sc}\,d)| \,e^{-\frac{d   }{\xi } }}{ \pi^2 \,d\, \Delta }\nn
 &\hspace*{2 cm}\times \int_0^{\pi/2}   d\theta' \,
 K_0\left( \frac{ |\alpha~ \text{sin}(\theta')| \Delta   } {\hbar\,u} \right), \label{SS'2}
 \end{align}
 where  $K_0( p)=\int_0^\infty dx \frac{\text{cos}(px)}{\sqrt{x^2+1}}$ is the modified Bessel function of the second kind.
We remark that for non-interacting systems with  $K_{1,3}(0)=1$, the derivatives of $\tilde{t}_{int}$, $\tilde{t}_{ext}$ are finite rather than zero as these dimensionless quantities depend explicitly on the cut-off $\alpha$ as $\tilde{t}_{int}, \tilde{t}_{ext} \propto \alpha^{3/2}$. However, the non-interacting case is reproduced correctly for physical quantities, $dt_{int}/{dl}$= $dt_{ext}/{dl}= 0$.
\begin{figure}[t]
 \includegraphics[width=0.36 \textwidth]{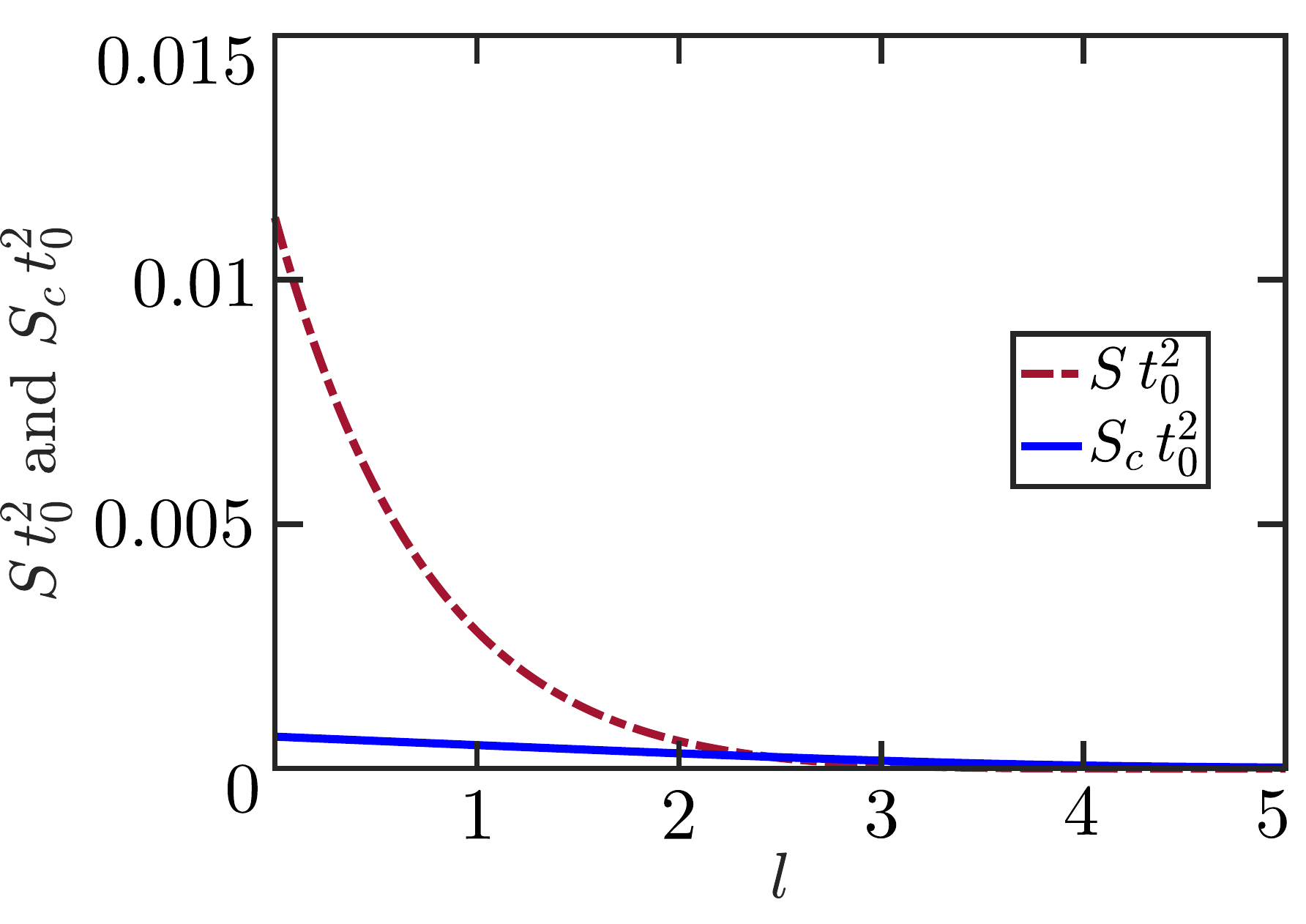}
 \caption{\label{sfunc} The source terms, $S \, t_0^2$ and $S_c \, t_0^2$ as a function of the RG flow parameter $l$, see Eqs. (\ref{SS'1}) and (\ref{SS'2}). Initially the interwire source term $S_c\, t_0^2$ (blue solid) is smaller compared to the intrawire source term $S\, t_0^2$ (red dashed), however, both vanish rapidly for large $l$. The used parameter values are $t_0=\tilde t_{int}(0)=\tilde t_{ext}(0)=3.8 \times 10^{-5}$,
$\Delta=0.35\, \text{meV}$,
$u=10^4~ \text{m/s}$,
$v_{F,sc}= 10^6~ \text{m/s}$, $\alpha_0=1 \,\text{nm}$, $d=15 \,\alpha_0$, $L= 1\, \mu \text{m}$, and  $\alpha_{sc}=1/k_{F,sc}=1 \buildrel _\circ \over {\mathrm{A}}$.  
}
\end{figure}

\begin{figure*}[t]
\begin{center} \begin{tabular}{ccc}
\epsfig{figure=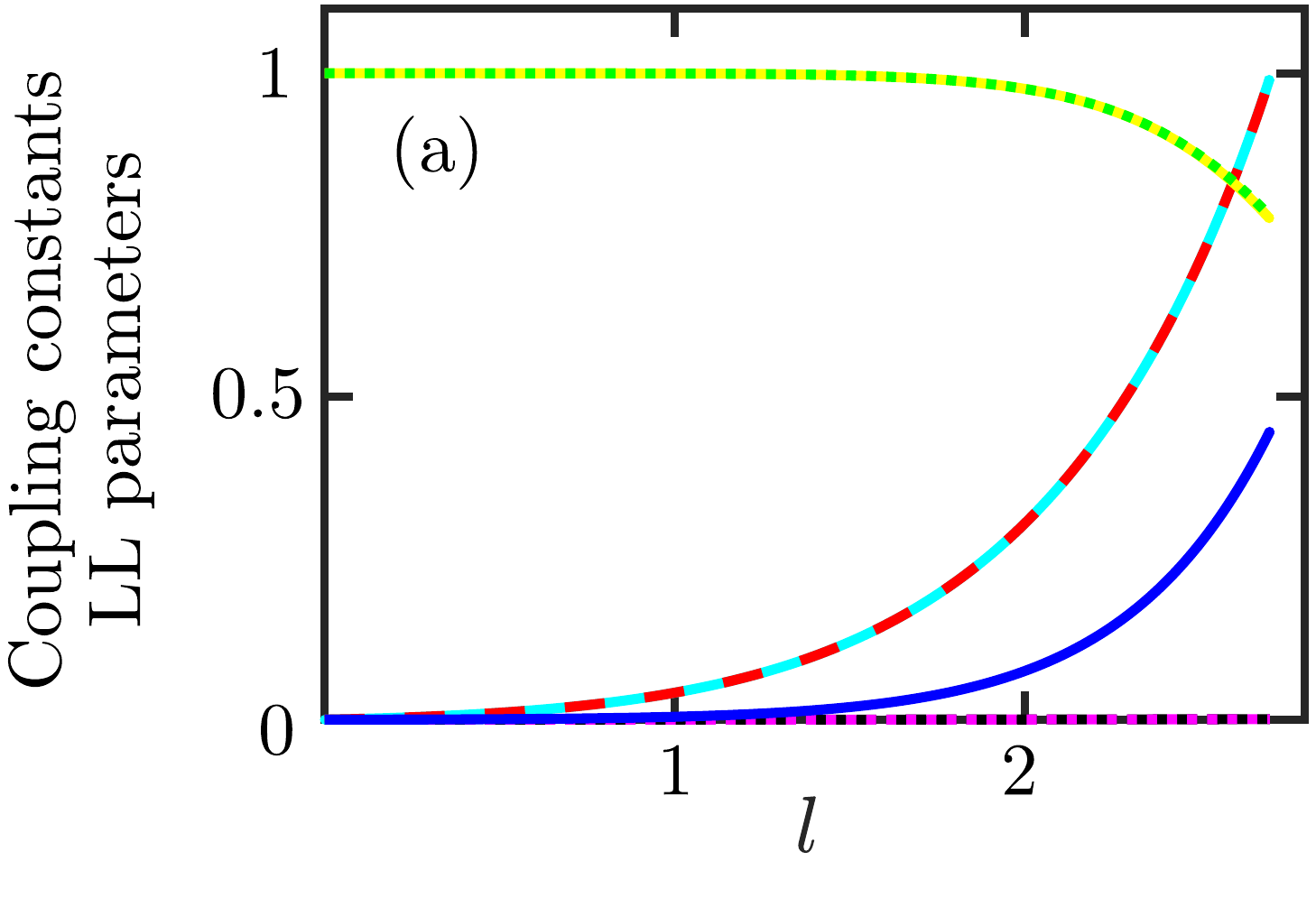,width=2.35in,height=1.6in,clip=true} &\hspace*{-0.cm}
\epsfig{figure=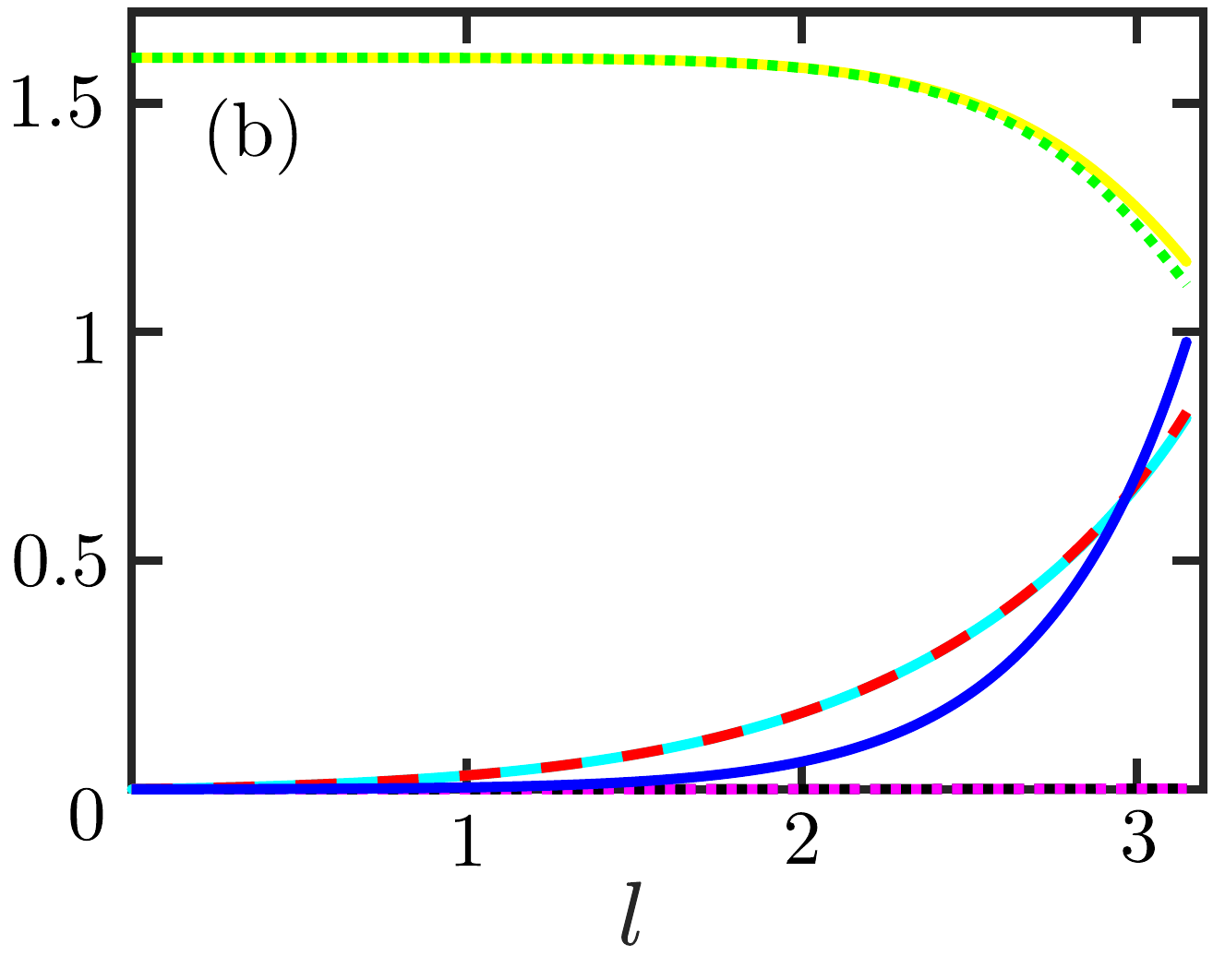,width=1.96in,height=1.59in,clip=true} &\hspace*{-0.cm}
\epsfig{figure=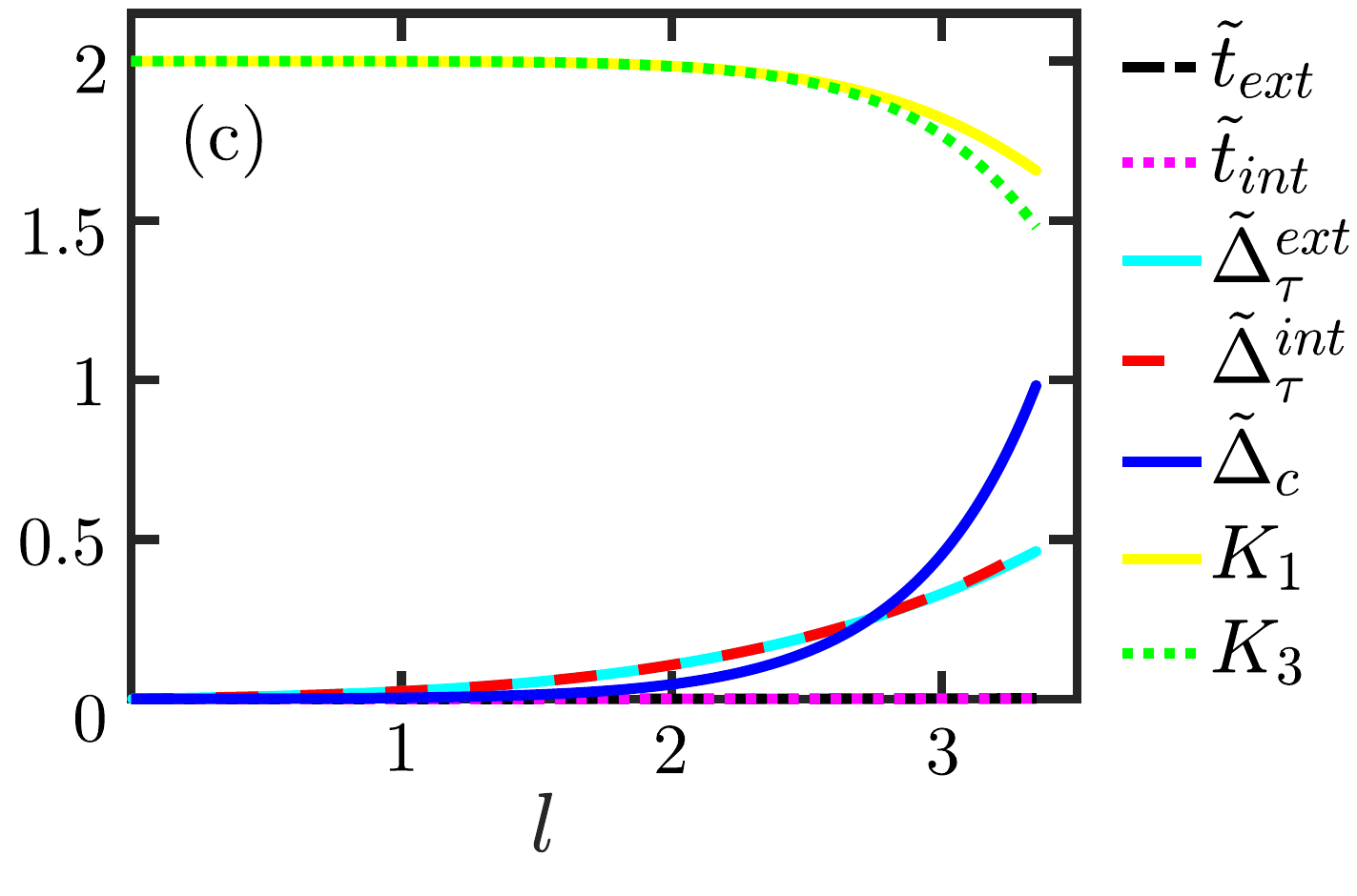,width=2.51in,height=1.59in,clip=true} 
\end{tabular} \end{center}
\caption{\label{fig_S_2} The RG flow of dimensionless coupling constants and LL parameters as a function of flow parameter $l$ obtained numerically in the source term approach from Eq. (\ref{rg4}). The initial values of LL parameters are chosen as (a) $K_{1,3}(0)=1$, (b) $K_{1,3}(0)=1.6$, and (c) $K_{1,3}(0)=2$. 
(a) For non-interacting systems, the intrawire pairing always dominates over the interwire pairing, $\tilde \Delta_\tau^{int}>\tilde \Delta_c$, resulting in the system being in a trivial phase. (b) Due to strong repulsive interactions, there is a crossover between intrawire pairing amplitude ($\tilde \Delta_\tau^{int}$) and interwire pairing amplitude ($\tilde \Delta_c$) at the end of the RG flow. (c) As the interaction strength is increased, $\tilde \Delta_c$ reaches the strong coupling limit much faster, which indicates that the RG flow brings the system into the topological phase.   In addition,  we calculate numerically the proximity-induced gaps in physical units: for (a), (b), and (c), the final flow values of these gaps are $\Delta_c/\Delta=0.33$, $0.81$, and $0.66$, and $\Delta_\tau^{int/ext}/\Delta=1$, $0.68$, and $0.33$, respectively. We note that we also stop the RG flow when a proximity gap reaches $\Delta$ ~\cite{footnote2}.  The other parameter values are fixed to 
$\Delta=0.35~ \text{meV}$,
$u=10^4~ \text{m/s}$,
$v_{F,sc}= 10^6~ \text{m/s}$, $\alpha_0=1\,\text{nm}$, $d=15\,\alpha_0$, $L=1 \,\mu \text{m}$,  and $\alpha_{\text{sc}}=1/k_{F,sc}=1 \buildrel _\circ \over {\mathrm{A}}$. We use the initial conditions: $\tilde t_{int,ext}(0)=3.8 \times 10^{-5}$ and $\tilde \Delta_\tau^{int,ext}(0)=\tilde\Delta_c(0)=0$.}
\end{figure*}
The source terms $S $ and $S_c$ are monotonically decreasing functions of the flow parameter $l$, see Fig. \ref{sfunc}. We further note that the crossed Andreev source term $S_c$ gets exponentially suppressed with increasing distance $d$ between the NWs on the scale of the coherence length $\xi$ of the SC (with power law correction $1/d$), and, moreover, oscillates on the scale of the Fermi wavelength of SC. Since this source term generates  the crossed Andreev pairing  in the RG flow, the same parameter dependence holds for $\Delta_c$
(possibly renormalized by interactions), which, again, is consistent with the non-interacting case obtained before~\cite{Reeg_a}.

Next, we solve the set of coupled RG  equations  [see Eq.~(\ref{rg4})]  numerically and plot the coupling constants and LL parameters as a function of flow parameter $l$, see Fig.~\ref{fig_S_2}. First, we explore how the flow depends on the initial values of the LL parameters $K_{1,3}(0)$ for fixed initial values of the source term $(S \,\tilde t^2_{int/ext})(l=0)$.  The initial conditions for the superconducting pairing amplitudes are $\tilde\Delta_\tau^{int}(0)=\tilde\Delta_\tau^{ext}(0)=\tilde\Delta_c(0)=0$, so only due to the presence of the source term proximity superconductivity arises.

Under the RG flow, the generated pairing amplitudes  $\tilde\Delta_\tau^{int/ext}$ and $\tilde\Delta_c$ become non-zero and grow.
In non-interacting systems, $K_{1,3}(0)=1$  as well as $S > S_c$, so the intrawire pairing amplitude is always greater than the interwire (crossed Andreev) pairing amplitude, which corresponds to the trivial phase of the system, see Fig. \ref{fig_S_2}\,(a). In contrast to that, in the presence of repulsive electron-electron interactions, described by the initial conditions $K_1(0)=K_3(0)>1$, at $l=0$ only the source terms are nonzero, thus at small values of $ l$, $\tilde\Delta_\tau^{int}>\tilde\Delta_c$, however as soon as the superconducting pairing amplitudes become finite, they also begin to influence the flow equations, see Eq. (\ref{rg4}). At large values of $l$, the flow equations are then governed by the pairing terms, which become  larger than the source terms.
In the later part of the flow, the interwire pairing will dominate over the intrawire one. 
In other words, the crossed Andreev  pairing ($\tilde\Delta_c$) should reach the strong coupling limit much faster than the direct   pairings ($\tilde\Delta_\tau^{int}$ and $\tilde\Delta_\tau^{ext}$). This eventually drives the system into the topological phase [see Figs. \ref{fig_S_2}\,(b) and (c)]. As in previous sections, we stop the RG flow whenever one of the coupling constants reaches unity. 

\begin{figure}[b!]
        \centering
                \includegraphics[width=0.38\textwidth]{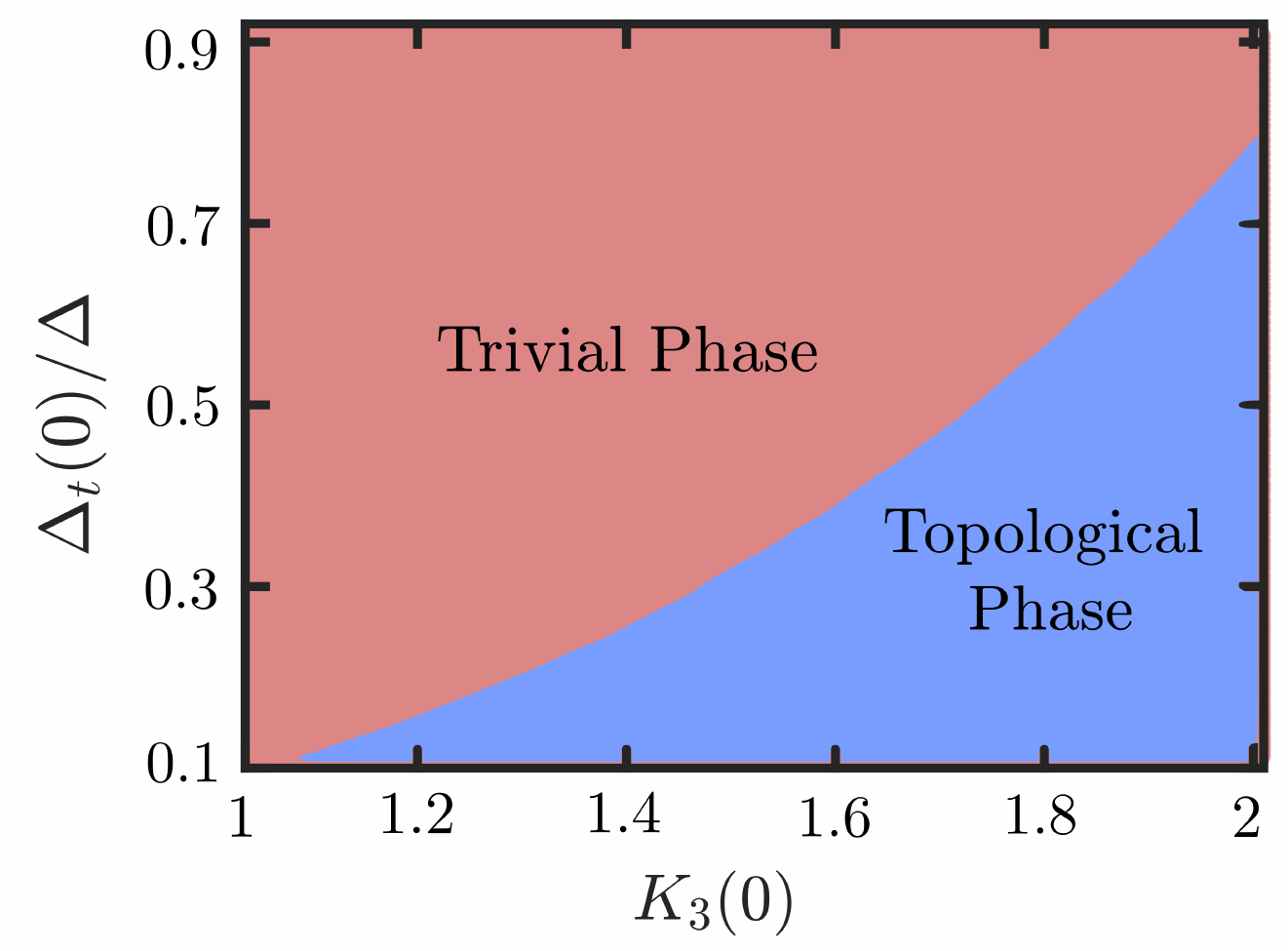} 
        \caption{\label{Phase_t} The phase diagram as a function of initial values of the source term $\Delta_t(0)/\Delta$  and LL parameter $K_3(0)$ obtained numerically by solving the RG equations with source terms, see Eq. (\ref{rg4}).
Here, $\Delta$ is the gap of the bulk SC and $\Delta_t(0)=S(0)\,\tilde t^2_{int/ext}(0) \frac{\hbar u}{\alpha_0}$ is the source term where we put back the $\hbar$ factor for proper energy units.  Again, whenever one of the coupling constants reaches unity, we stop the flow.  If the crossed Andreev pairing amplitude $\Delta_c (l)> \Delta_\tau^{int}(l)$ dominates, the system is in the topological phase (blue area). If the direct pairing amplitude wins, the system is in the trivial phase (red area). For small initial values of the tunneling amplitude, the system tends to be always in the topological phase. If the tunneling is increased or the distance $d$ between the NW grows,  stronger and stronger interactions are required to bring the system into the topological phase.
 The initial conditions are $\tilde \Delta_\tau^{int}(0)=\tilde\Delta_\tau^{ext}(0)=\tilde \Delta_c(0)=0$ and $K_1(0)=K_3(0)$, and the parameter values are $\Delta=0.35~ \text{meV}$, $u=10^4~ \text{m/s}$, $v_{F,sc}= 10^6~ \text{m/s}$, $\alpha_0=1\,\text{nm}$, $d=15\,\alpha_0$, 
 $L=1 \,\mu \text{m}$, and $\alpha_{\text{sc}}=1/k_{F,sc}=1 \buildrel _\circ \over {\mathrm{A}}$.  
         }
\end{figure}

In addition, we explore how the phase diagram depends on different initial conditions. In particular, on the tunneling strengths, $\tilde t^2_{int/ext}(0)$, and LL parameters, $K_1$ and $K_3$, see Fig.~\ref{Phase_t}.  The energy scale for the tunneling terms is defined via the source term as $\Delta_t=S(0)\,\tilde t^2_{int/ext}(0) \frac{\hbar u}{\alpha_0}$.
Similarly, to the previous section, as the ratio between the initial value of source term and superconducting gap increases, we need stronger and stronger interactions to reach the topological phase in the system. This can be understood in the following sense. The initial values of source term are always favor the intrawire pairing as $S>S_c$, see Fig. \ref{sfunc}. Only at large $l$, the crossed Andreev pairing grows and begins to dominate. However, if the intrawire pairing was large from the beginning, the intrawire pairing has already reached the strong coupling regime and the flow must be stopped~\cite{footnote2}, see Fig. \ref{fig_S_2}. As a result, the crossed Andreev term does not have a chance to develop.

To conclude, we note that the two phase diagrams obtained by solving the RG equations for the effective Hamiltonian (see Fig. \ref{rgplot1}) and by using the source term approach (see Fig. \ref{fig_S_2}) qualitatively look similarly. However, quantitatively they are different in the following way: (1) The crossed Andreev pairing amplitude $\tilde \Delta_c$ reaches the strong coupling limit faster in Fig. \ref{fig_S_2}(c) compared to Fig. \ref{rgplot1}. (2) As shown in Fig. \ref{fig_S_2}, at the beginning of the RG flow, the contribution coming  from the intrawire source term $S$ is always greater than the one from the interwire term, $S_c$ (see Fig. \ref{sfunc}), which results in $\tilde \Delta_\tau^{int/ext}$ being greater than $\tilde \Delta_c$. In contrast to that, for larger values of flow parameter $l$ and due to strong repulsive electron-electron interactions in the NWs, there is a crossover between  $\tilde \Delta_\tau^{int/ext}$ and $\tilde \Delta_c$. 
However, as seen from Fig. \ref{rgplot1},  if $\tilde \Delta_\tau^{int}(0)=\tilde \Delta_c(0)$, $\tilde \Delta_c$ is always greater than $\tilde \Delta_\tau^{int/ext}$ for repulsive  interactions in the NWs, while 
for initial values $\tilde \Delta_\tau^{int}(0)>\tilde \Delta_c(0)$, there is also a crossover between $\tilde \Delta_\tau^{int/ext}$ and $\tilde \Delta_c$, and the topological phase can be reached.
Thus, the results obtained in Sec. \ref{rg} also hold  if we start from a more microscopic approach in terms of a tunneling Hamiltonian between superconductor and NWs.  Finally, we note that we have also checked numerically the RG flows for disordered NWs in the source term approach,  and we got essentially the same qualitative results as already presented  in Sec. \ref{rg}.  Thus, we do not consider the RG flow equation of disorder and backscattering terms in this section.

However,  there is a quantitative difference between the two approaches in that stronger electron-electron interactions (larger $K_3$) are needed in the tunneling approach in order
to reach the topological phase. This can be understood by the following qualitative reasoning (see also Refs. \onlinecite{Recher_Sukhorukov,DL1}).
In the tunneling approach, the suppression of the direct pairing is  less pronounced than in the phenomenological approach. This is so because the two electrons from a Cooper pair enter the NW in a second order tunneling process,  which implies that  the electrons of the NW interact with  each tunneling electron (more or less) one by one  since they hop from the SC on the NW one after the other in a co-tunneling fashion, with some virtual delay time between them. This delay time is roughly inversely proportional to the SC gap $\Delta$-the energy cost of the virtual excitation on the SC (ignoring correlation effects in the NW). Thus, the smaller $\Delta$ the more the electrons are separated in time, and the less difference
we get between direct and crossed Andreev processes.
In contrast, in the phenomenological model, the electrons are added as a whole Cooper pair (with twice the electron charge) to the NWs, 
which gives rise to stronger repulsive interactions with the electrons in the NWs.
 As a consequence,  the direct and the crossed Andreev  processes  are less distinguishable by the interactions in the tunneling  than in the phenomenological approach and it takes stronger interactions in the former case to make the crossed Andreev process to dominate over the direct one.

To summarize, in the microscopic source-term approach, the interactions are taken into account at a more fundamental level than in the effective Hamiltonian approach. 
In particular, interactions modify already the  tunneling process that generates the superconducting pairing terms in the NWs, while in the effective Hamiltonian approach we add interactions only at a later stage after the pairing gap is already formed. However, very similar conclusions are reached in both approaches.

\section{Conclusions and Outlook}
\label{con}
In this work, we  studied a setup consisting of two Rashba NWs coupled to a bulk three-dimensional $s$-wave superconductor. We focused on the interplay between direct (intrawire) and crossed Andreev (interwire) superconducting pairing processes. Standard bosonization techniques were used to treat strong electron-electron interactions and weak uncorrelated Gaussian disorder. For the latter, we employed the replica trick of disorder averaging.  We performed an RG analysis to determine which terms dominate and identified the parameter regime for which the system is in the topological phase, with a KMBS at each end of the double-NW system. 
In particular, the crossed Andreev pairing amplitude $\tilde \Delta_c$ reaches the strong coupling limit for $K_{1,3}>1$, while for $K_{1,3}<1$ it flows to zero as $\tilde \Delta_c=0$ is a stable fixed point. The value of spin and charge LL constants in the NWs should satisfy the condition $K_{\tau s}>K_{\tau c}$,
which is possible only for  repulsive electron-electron interactions in the NWs. By evaluating numerically the phase diagram, we have confirmed that the topological phase is stable against weak disorder.
Generally, electron-electron interactions lower the value of all types of gaps in the NWs\cite{Suhas}, however, the interwire (crossed Andreev) pairing amplitude gets reduced less than the intrawire (direct) pairing amplitude, enabling eventually the topological phase hosting KMBSs. Importantly, the topological phase is achieved even if the system is initially in the trivial phase with dominant direct superconducting pairing as predicted by non-interacting theories\cite{Reeg_a}.

We have reached essentially the same conclusions in two independent approaches. In the effective Hamiltonian approach, superconducting pairings are explicitly included in the Hamiltonian. In the more microscopic approach, the source term, arising from the tunneling between NWs and bulk SC, is responsible for generating superconducting correlations in the NWs. Apart from minor quantitative differences between the two approaches, 
both show that strong electron-electron interactions enable the topological phase even if the system is initially (without interactions) in the trivial phase and even in the presence of moderate disorder. Thus, the double-NW system discussed in this work is a promising candidate for observing Majorana fermions in the absence of magnetic fields.

In the present work, we have focused on the topological phase hosting Majorana fermions. However, the two RG approaches developed here can also be applied to fractional topological phases, hosting parafermions or fractional Majorana fermions\cite{Oreg_para,JK6},  which is, however, beyond the scope of the present work. Moreover, our findings can also be extended straightforwardly from NWs to one-dimensional helical edges of two-dimensional topological insulators\cite{Molenkamp}. In the approach presented here the interior and exterior branches were treated independently. Thus, the characteristic behavior of superconducting pairings induced in helical edge states can be mapped to our model by retaining only terms acting on the interior branches of the spectrum. Again, one would expect that the crossed Andreev pairing dominates in the regime of strong electron-electron interactions and is stable against weak disorder treated as in Ref. [\onlinecite{JK7}].

Alternatively, crossed Andreev  pairing also plays an important role in many proposals for parafermions in quantum Hall systems coupled to bulk $s$-wave superconductors. Based on our analysis, we can expect that strong electron-electron interactions   will suppress   the proximity-induced pairing gap in the chiral edge channels. However, in this case, the  RG analysis should be carefully redone by including time-reversal symmetry breaking terms to take  Zeeman splittings and orbital magnetic effects properly into account. For such a treatment, the coupled-wire model seems to be a most suitable starting point as it will allow one to describe both integer and  fractional filling factors \cite{Lederer,Gorkov,Lubensky,JK8,Teo,JK9,JK10,JK11,Neupert,Sagi,JK12,JK13}. 
The source term approach also opens up the possibility to obtain the dependence of crossed-Andreev pairings on the distance between two NWs. In our calculations, we worked with an effectively infinite bulk $s$-wave superconductor. As a consequence, proximity-induced pairing terms are independent of the  size of the SC. However, it would be interesting to consider SCs of finite geometry such as thin films, as done for non-interacting systems \cite{JK14}. In this case, one needs to use appropriate Green functions which account for the finite-size effects of the SC.  We believe that similar progress can be made along the lines shown here.

\acknowledgements

We  acknowledge useful discussions with P. P. Aseev and C. Reeg.  This work was supported by the Swiss National Science Foundation (SNSF) and NCCR QSIT.

\onecolumngrid
\appendix
\newpage

\section{Transformation relations of LL parameters} \label{alp}

In this Appendix, we compute relations between the new bosonic field velocities $u_i$ and LL parameters $ K_i$ ($i=1,2,3,4$) and the charge-spin velocities $u_{\tau c}$, $u_{\tau,s} $ and LL parameters  $K_{\tau c}$, $K_{\tau,s} $ for each $\tau$-NW. First, we define the charge-spin bosonic fields as $(\phi_{\tau,c}, \phi_{\tau,s})$ and their conjugate fields as ($\theta_{\tau,c}$, $\theta_{\tau,s}$). These fields obey the commutation relation $[\phi_{\tau,c/s}(x),\theta_{\tau',c/s}(x')]= i \pi\, \delta_{\tau \tau'} \text{sgn}(x'-x)/2$. 
Using Eq. (\ref{chiralfield}), the Hamiltonian $H_0$ given by Eq.~(\ref{H0}) in the main text takes the following form:
\begin{align}
&H_0= \sum_{\tau}  \int \frac{dx}{2\pi} \Bigg[
u_{\tau,c} \Big[\frac{\left ( \partial_x \phi_{\tau,c}  \right )^2 } {K_{\tau,c}} + K_{\tau,c} \left ( \partial_x \theta_{\tau,c}  \right )^2\Big]+ u_{\tau,s} 
\Big[\frac{\left ( \partial_x \phi_{\tau,s}  \right )^2 } {K_{\tau,s}} + K_{\tau,s} \left ( \partial_x \theta_{\tau,s}  \right )^2\Big]\,\Bigg].
\label{AH0}
\end{align}
Next, we change to the new bosonic field basis ($\phi_j$ and $\theta_j$, where $j=1,2,3,4$) introduced in Eq. (\ref{nfields}) of the main text with commutation relations defined as $[\phi_j(x), \theta_{j'}(x')]=i \pi\, \delta_{j,j'} \text{sgn}(x'-x)/2$.
As a result,  Eq. (\ref{AH0}) takes the form
\begin{align}
H_0=   \int  \frac{dx}{2\pi} \Bigg[&
\frac{u_{1,c}}{2} \Big\{\frac{\left ( \partial_x (\theta_1+\theta_3)  \right )^2 } {K_{1,c}} + K_{1,c} \left ( \partial_x (\phi_1+\phi_3)  \right )^2\Big\}+ \frac{u_{1,s}}{2} 
\Big\{\frac{\left ( \partial_x (\phi_3-\phi_1)  \right )^2 } {K_{1,s}} + K_{1,s} \left ( \partial_x (\theta_3-\theta_1)  \right )^2\Big\}\,\nn
&+\frac{u_{\bar 1,c}}{2} \Big\{\frac{\left ( \partial_x (\theta_2+\theta_4)  \right )^2 } {K_{\bar 1,c}} + K_{\bar 1,c} \left ( \partial_x (\phi_2+\phi_4)  \right )^2\Big\}+ \frac{u_{\bar 1,s}}{2} 
\Big\{\frac{\left ( \partial_x (\phi_4-\phi_2)  \right )^2 } {K_{\bar 1,s}} + K_{\bar 1,s} \left ( \partial_x (\theta_4-\theta_2)  \right )^2\Big\}\,
\Bigg].
\label{AH1}
\end{align}
We recall that $u_i$ and  $K_i$ are the  new velocities and LL parameters, respectively. We consider only the diagonal terms in Eq. (\ref{AH1}) as the non-diagonal terms are marginal operators which are negligible under the RG flow \cite{Braunecker}. The Hamiltonian $H_0$ in Eq. (\ref{AH1}) takes the form 
\begin{align} 
&H_0= \sum_{i=1,2,3,4} u_{i}  \int  \frac{dx}{2\pi}
\Big[\frac{\left ( \partial_x \phi_{i}  \right )^2 } {K_{i}} + K_{i} \left ( \partial_x \theta_{i}  \right )^2\Big]\,,
\label{aH0}
\end{align}
with the following constraints
\begin{align}
\frac{u_1} {K_1}&=\frac{u_3} {K_3}= \frac{u_{1,s}}{2 K_{1,s}}+\frac{K_{1,c} u_{1,c}}{2},~~
\frac{u_2} {K_2}=\frac{u_4} {K_4}= \frac{u_{\bar 1,s}}{2 K_{\bar 1,s}}+\frac{K_{\bar 1,c} u_{\bar 1,c}}{2},\nn
u_1 K_1&=u_3 K_3= \frac{u_{1,c}}{2 K_{1,c}}+\frac{K_{1,s} u_{1,s}}{2},~~u_2 K_2=u_4 K_4= \frac{u_{\bar 1,c}}{2 K_{\bar 1,c}}+\frac{K_{\bar 1,s} u_{\bar 1,s}}{2}.
\label{AH2}
\end{align}
This allows us to determine $u_i$ and $K_i$,
\begin{align}
u_{1/2}=&u_{3/4}= \frac{1}{2} \sqrt{ \left( \frac{u_{1/\bar 1c}}{K_{1/\bar 1c}} + K_{1/\bar 1s}\, u_{1/\bar 1s}\right)\left( \frac{u_{1/\bar 1s}}{K_{1/\bar 1s}} + K_{1/\bar 1c}\, u_{1/\bar 1c}\right)}\,, \nn \\ 
K_{1/2}=&K_{3/4}= \sqrt{\frac{\frac{u_{1/\bar 1c}}{K_{1/\bar 1c}} + K_ {1/\bar 1s} \,u_{1/\bar 1s}}{\frac{u_{1/\bar 1s}}{K_{1/\bar 1s}} + K_{1/\bar 1c}\, u_{1/\bar 1c}}}\,.
\end{align}
For an  ideal LL (Galilean invariant continuum model) we have $u_{\tau, c/s}=\frac{v_{F,\tau}}{K_{\tau, c/s}}$. This leads us to the expressions  given in Eqs. (\ref{u}) and (\ref{K}) of the main text.

\section{Operator product expansion (OPE) }
\label{aope}
In this Appendix, we write the expressions for OPE \cite{Cardy,Schoeller,Senechal,Tsuchiizu} used later in App. \ref{arg} to derive the RG equations. The kinetic part of the Hamiltonian is given by
\begin{align}
&H_{0a}= u  \int \frac{dx}{2\pi} 
\Big[\frac{\left ( \partial_x \phi  \right )^2 } {K} + K \left ( \partial_x \theta  \right )^2\Big]\,,
\label{H0a}
\end{align}
where $K$ is the LL parameter and $\phi$ and its conjugate field $\theta$ are bosonic fields with the only nonzero commutation relation given by $[\phi(x), \theta(x')]=i \pi\, \text{sgn}(x'-x)/2$.
We define the complex coordinates $(z,\bar z)$ as $z=-i\,x+u\,t$ and $\bar z=i\,x+u\,t $, where $x$ and $t$ are position and imaginary time coordinates, respectively. The corresponding derivatives are given by  $\partial_z=-\frac{1}{2}\Big( \frac{\partial_t}{u}-i\,\partial_x\Big)$ and $\partial_{\bar z}=-\frac{1}{2}\Big( \frac{\partial_t}{u}+i\,\partial_x\Big)$\cite{Senechal,TG}. 
In the OPE expressions, we use the following relations for bosonic operators $A$ and $B$~\cite{TG,Shankar}, 
\begin{align}
&e^{A}\ e^{B}=:e^{A+B}: e^{\langle AB +\frac{A^2+B^2}{2}\rangle_0}\, , 
\label{exp1} 
 \end{align}
where $:C:$ denotes normal ordering of the operator $C$. In what follows, we will be also using the following expressions:
\begin{align}
&\langle [\phi(z,\bar z)- \phi(0,0)]^2\rangle_0 = K \,\text{ln}\frac{|z|}{\alpha}\,, \,\,\,\,\,\,\,\,\, \langle [\phi(0,0)]^2\rangle_0 \sim -\frac{K}{2} \,\text{ln} \alpha\, ,  \nonumber \\
&\langle [\theta(z,\bar z)- \theta(0,0)]^2\rangle_0 =\frac{1}{K} \,\text{ln}\frac{|z|}{\alpha}\,,\,\,\,\,\,\,\,\,\, \langle [\theta(0,0)]^2\rangle_0 \sim -\frac{1}{2K} \,\text{ln} \alpha\,,
\label{exp11} 
 \end{align}
where the expectation value $\langle \cdots\rangle_0$ is taken with respect to the  LL Hamiltonian $H_{0a}$ defined in Eq. (\ref{H0a}), and where ``$\sim$" indicates omission of constant units. 
 
 Further, we write the OPEs\cite{Cardy,Schoeller,Recher,Senechal,Tsuchiizu} for the conjugate $\phi$ and $\theta$ fields, 
\begin{align}
&e^{i \lambda \phi(z,\bar z)}e^{-i \lambda \phi(0,0)}= \frac{1}{(|z|/\alpha)^{\lambda^2 K/2}}+\frac{\lambda }{(|z|/\alpha)^{\lambda^2 K/2}}(z J_\phi - \bar z \bar J_\phi)+\frac{i\,\lambda \, \alpha^2}{(|z|/\alpha)^{\lambda^2 K/2-2}}(\partial_z \partial_{\bar z})\phi\nn
 &\hspace{35pt}+\frac{i \lambda}{2 (|z|/\alpha)^{\lambda^2 K/2}}[z^{2} (\partial^2_z \phi)+ \bar z^2 (\partial^2_{\bar z} \phi)] \nn 
&\hspace{35pt}+ \frac{ \lambda^2 }{2 (|z|/\alpha)^{\lambda^2 K/2}}[z^2 :J^2_{\phi}:+ \bar z^2 :\bar J^2_{\phi}:]- \frac{\lambda^2 \, \alpha^2 }{(|z|/\alpha)^{\lambda^2 K/2 -2}} J_\phi \bar J_\phi+\cdots, \\
&e^{i \lambda \theta(z,\bar z)}\,e^{-i \lambda \theta(0,0)}= \frac{1}{(|z|/\alpha)^{\lambda^2/ 2K}}+\frac{\lambda}{(|z|/\alpha)^{\lambda^2/2K}}(z J_\theta - \bar z \bar J_\theta)+\frac{i\,\lambda\, \alpha^2}{(|z|/\alpha)^{\lambda^2 /2K-2}}(\partial_z \partial_{\bar z})\theta\nn
&\hspace{35pt}+\frac{i \lambda}{2 (|z|/\alpha)^{\lambda^2/2K}}[z^2 (\partial^2_z \theta)+ \bar z^2 (\partial^2_{\bar z} \theta)]+ \frac{ \lambda^2}{2 (|z|/\alpha)^{\lambda^2/2 K}}[z^2 :J^2_{\theta}:+ \bar z^2 :\bar J^2_{\theta}:]\nn
&\hspace{35pt}- \frac{\lambda^2\, \alpha^2}{(|z|/\alpha)^{\lambda^2/2K -2}} J_\theta \bar J_\theta+\cdots,\\
&e^{i \lambda [\phi(z,\bar z)+\theta(z,\bar z)]}\,e^{-i \lambda [\phi(0,0)+\theta(0,0)]}=  \Big[\frac{1}{(|z|/\alpha)^{\lambda^2(K+1/K)/ 2}}+\frac{\lambda}{(|z|/\alpha)^{\lambda^2(K+1/K)/2}}[z (J_\phi+J_\theta) - \bar z (  \bar J_\phi +\bar J_\theta)]\nn
&\hspace{35pt} +\frac{i\lambda \, \alpha^2}{(|z|/\alpha)^{\lambda^2 (K+1/K)/2-2}}(\partial_z \partial_{\bar z})(\phi+\theta)
+\frac{i \lambda}{2 (|z|/\alpha)^{\lambda^2(K+1/K)/2}}(z^2 \partial^2_z + \bar z^2 \partial^2_{\bar z})(\phi+\theta) \nn
&\hspace{35pt}+ \frac{ \lambda^2}{2 (|z|/\alpha)^{\lambda^2 (K+1/K)/2}}[z^2 ( :J^2_{\phi}:+:J^2_{\theta}:)+ \bar z^2 (:\bar J^2_{\phi}:+:\bar J^2_{\theta}:)]\nn
&\hspace{35pt}- \frac{\lambda^2\, \alpha^2}{(|z|/\alpha)^{\lambda^2(K+1/K)/2 -2}} ( J_\phi+J_\theta) ( \bar J_\phi +\bar J_\theta)+\cdots \Big] e^{\lambda^2\{\langle \phi(z,\bar z) \theta(0,0)+ \theta(z,\bar z) \phi(0,0)\rangle_0\}},
\label{dc} 
 \end{align}
where $\lambda$ is a real constant and  $J_\phi=[i \partial_z\phi(z,\bar z)]|_{(z,\bar z) = (0,0)}, \, \bar J_\phi=[-i \partial_{\bar z}\phi(z,\bar z)]|_{(z,\bar z) = (0,0)}$; $J_\theta(z)=[i \partial_z \theta(z,\bar z)]|_{(z,\bar z) = (0,0)}$, and $\bar J_\theta(\bar z)=[-i \partial_{\bar z}\theta(z,\bar z)]|_{(z,\bar z) = (0,0)}$. In the above OPEs, the terms that renormalize the kinetic energy terms are given by
\begin{align}
 J_\phi \bar J_\phi&= \partial_z\phi \, \partial_{\bar z}\phi= \frac{(\partial_x \phi)^2+(\partial_t \phi)^2/u^2}{4}, \nn 
J_\theta \bar J_\theta&=\partial_z\theta \partial_{\bar z}\theta=\frac{(\partial_x \theta)^2+(\partial_t \theta)^2/u^2}{4}.
\label{kineticterms}
\end{align}

\section{Derivation of RG equations in the effective Hamiltonian approach}
\label{arg}

In this Appendix, we derive the RG flow equations for coupling constants and LL parameters in the effective Hamiltonian approach. The effective Hamiltonian is defined in Eq. (\ref{Heff}) of the main text. Before calculating the RG flow equations for our involved Hamiltonian, we show the basic steps how to perform the RG analysis for the simple Hamiltonian, $H=H_{0a} + \frac{\Lambda}{\pi \alpha} \int dx \cos(\lambda \,\phi)=H_{0a} + \frac{\tilde \Lambda \, u}{\pi \alpha^2} \int dx \cos(\lambda \,\phi)$, where $H_{0a}$ is defined in Eq. (\ref{H0a}). Here, the following symbols were introduced: $\Lambda$ (coupling constant with dimension of energy), $\tilde \Lambda =\Lambda  \alpha/u $ (dimensionless coupling constant), $\alpha$  (lattice constant), $\lambda$ (real constant), and $u$ (Fermi velocity in the NWs). Before obtaining the  RG flow equations for $\tilde \Lambda$ and $K$, we write down the OPE for $\cos(\lambda\phi)$ for $(z_{1/2},\bar z_{1/2}) \rightarrow (z_c,\bar z_c)$, where $z_c=(z_1+z_2)/2$ is the center-of-mass coordinate. In what follows, we will keep only singular terms, which leads us to 
\begin{align}
\cos[\lambda\,\phi(z_1,\bar z_1)]\, \cos[\lambda \,\phi(z_2,\bar z_2)]&= [e^{i \lambda \,\phi(z_1,\bar z_1)} e^{-i \lambda \,\phi(z_2, \bar  z_2)}+e^{-i \lambda \,\phi(z_1,\bar z_1)} e^{i \lambda \,\phi(z_2, \bar  z_2)}]/4 , \nn
&= \frac{1}{2(|z_1-z_2|/\alpha)^{\lambda^2 K/2}}-\frac{\lambda^2 \alpha^2}{2(|z_1-z_2|/\alpha)^{\lambda^2 K/2 -2}}  [J_{\phi} \bar J_{\phi}]_{(z_c,\bar z_c)}+\cdots,
\label{ed1}
\end{align}
To obtain the RG flow equations,  the partition function is expanded in powers of the cosine term, which gives up to second order
\begin{align}
Z_a= Z_{0a} \,\Big\langle\, 1-  \frac{\tilde\Lambda u}{ \pi \, \alpha^2}\int dx \,dt\ \cos[\lambda \,\phi(x,t)]  + \frac{\tilde\Lambda^2 u^2}{2\, \pi^2 \, \alpha^4}\int dx_1\, dx_2 \,dt_1 \,dt_2\, \cos[\lambda \,\phi(x_1,t_1)] \cos[\lambda \,\phi(x_2,t_2)]  + \cdots  \Big\rangle_0 \,,
\label{Za}\end{align} 
where $Z_{0a}$ is the partition function for fixed point Hamiltonian $H_{0a}$. To implement the RG procedure, we change the cutoff from $\alpha$ to $\alpha +d\alpha$ and calculate the corresponding change in $\tilde \Lambda$ in such a way that the partition function is preserved\cite{Cardy}. First, we consider the second term in Eq. (\ref{Za}) and calculate the change in $\tilde \Lambda$. From Eq. (\ref{ed1}), the scaling dimension of $\cos(\lambda \, \phi)$ is half the power of $1/|z|$, {\it i.e.}, $\lambda^2 K/4$. Thus we obtain the RG flow equation for $\tilde \Lambda$ as
\begin{align}
\frac{d\tilde\Lambda}{dl}=\left( 2-\frac{\lambda^2 K}{4} \right)\tilde\Lambda, ~~~\text{where} ~dl=\frac{d\alpha}{\alpha}.
\label{ed2}
\end{align}
Next we consider the third term in Eq. (\ref{Za}) and again change the cutoff from $\alpha$ to $\alpha +d\alpha$. For obtaining the contribution from this term to the quadratic part of the Hamiltonian, we change to the center-of-mass coordinates, $X=(x_1+x_2)/2$, $T=(t_1+t_2)/2$, $x=x_1-x_2$ and $t=t_1-t_2$, which in terms of the complex coordinates are defined as $z_{1/2}=-i \,x_{1/2} + u \,t_{1/2}$, and take the form $z_c= (z_1+z_2)/2$ and $z=z_1-z_2$. We then change to polar coordinates $(r,\theta')$ with $\int dx (u\,dt)= \int r \, dr\, d\theta'= 2\pi \int r \,dr$ and split the integral over $r$ into two parts such that $\int_{r>(\alpha+d\alpha)} = \int_{r>\alpha}-\int_{\alpha}^{\alpha+d \alpha}$. The first integral contributes towards the original integral in the partition function and we only need to compute the integral  within $\alpha < r < \alpha + d\alpha$. We use the OPE given by Eq. (\ref{ed1}), where we focus on the second term, which  gives the renormalization of the LL parameter $K$. Thus, the contribution from the third term in the partition function which renormalizes $K$ is given by
\begin{align}
I^{d\alpha}= &-\int dx \, dt \,dX \,dT \,\, \frac{ \tilde \Lambda^2 u^2}{2\,\pi^2\, \alpha^4} \,\frac{\lambda^2 \, \alpha^2}{2\,|z/\alpha|^{\lambda^2 K/2 -2}}  \,[J_{\phi} \bar J_{\phi}]_{(X,T)}\nn
 =&-\frac{1}{16}\int dx \, dt \,dX \,dT \,\, \frac{\tilde \Lambda^2 u^2}{\pi^2 \, \alpha^2} \,\frac{\lambda^2 }{|z/\alpha|^{\lambda^2 K/2 -2}}\, \Big[(\partial_X \phi)^2+\frac{(\partial_T \phi)^2}{u^2}\Big].
 \label{ed3}
\end{align}
We change the $(x,t)$ to $(r,\theta')$ as described above and compute the integral within $\alpha < r < \alpha + d\alpha$.  Thus, using $\int_\alpha^{\alpha+d\alpha} f(r) dr= f(\alpha) d\alpha$, Eq. (\ref{ed3}) takes the form
\begin{align}
I^{d\alpha}=&\frac{\lambda^2 \tilde \Lambda^2 }{4} \frac{d\alpha}{\alpha} \int    \frac{\,dX \,dT \, }{2 \pi} \Big[u\,(\partial_X \phi)^2+\frac{(\partial_T \phi)^2}{u}\Big].
 \label{ed4}
\end{align}
Hence, we get from Eq. (\ref{ed4}) that, in order to preserve the partition function ($Z_a$), the LL parameter $K$ has to change in the following way
 \begin{align}
\frac{dK^{-1}}{dl}=\frac{\lambda^2 \tilde \Lambda^2}{4}\,\,\,\, \Rightarrow \,\,\,\,\frac{dK}{dl}=\frac{-(\lambda\, K \, \tilde \Lambda)^2}{4}.
\label{ed5}
\end{align}

Based on the discussion above, we explicitly derive the RG flow equations for the two terms from the Hamiltonian defined in Eq.~(\ref{Heff}) of the main text: for the superconducting pairing for exterior branches with the coupling amplitude $\Delta_1^{ext}$ and for crossed Andreev superconducting pairing  with the coupling amplitude $\Delta_c$. Note that we do not consider any crossterm between different coupling constants as they are less relevant compared to the original cosine terms\cite{TG}. The RG equations for the remaining coupling constants and LL parameters are determined by following the same procedure from Eqs. (\ref{ed1})-(\ref{ed5}).

First we consider the term $\frac{\Delta_1^{ext}}{\pi \alpha} \int dx \cos(2 \phi_1)=\frac{\tilde \Delta_1^{ext} u}{\pi \alpha^2} \int dx \cos(2 \phi_1)$. We put $\lambda=2$ and $K=K_1$ in the RG equations calculated in Eqs. (\ref{ed2}) and (\ref{ed5}), thus $\frac{d \tilde \Delta_1^{ext}}{dl}=\left( 2-K_1 \right)\tilde \Delta_1^{ext}$ and $\frac{dK_1}{dl}= -(\tilde \Delta_1^{ext})^2 K_1^2 $.

To calculate the OPE for the $\Delta_c$-term, $\frac{\tilde \Delta_c u}{\pi \alpha^2} \int dx ~[\cos( \phi_3+\phi_4+\theta_3-\theta_4)+\cos(\phi_3+\phi_4-\theta_3+\theta_4)]$, we need to generalize the previous procedure. The only nonzero commutation relation between the fields $\phi_3$, $\theta_3$, $\phi_4$, and $\theta_4$  is given by $[\phi_{3/4}(x),\theta_{3/4}(x')]= i\pi\, \text{sgn}(x'-x)/2$. 
We again consider only the most singular terms in the limit $(z_{1/2},\bar z_{1/2}) \rightarrow (z_c, \bar z_c)$ and write down the OPEs relevant for the $\tilde \Delta_c$ term,
\begin{align}
\cos[\phi_3(z_1,\bar z_1)+\phi_4(z_1,\bar z_1)&+\theta_3(z_1,\bar z_1)-\theta_4(z_1,\bar z_1)] \cos[\phi_3(z_2,\bar z_2)+\phi_4(z_2,\bar z_2)+\theta_3(z_2,\bar z_2)-\theta_4(z_2,\bar z_2)] \nn 
&= \big[e^{i (\phi_3(z_1,\bar z)+\phi_4(z_1,\bar z_1)+\theta_3(z_1,\bar z_1)-\theta_4(z_1,\bar z_1))}e^{-i (\phi_3(z_2,\bar z_2)+\phi_4(z_2,\bar z_2)+\theta_3(z_2,\bar z_2)-\theta_4(z_2,\bar z_2))}\nn 
&~~~~~~+e^{-i (\phi_3(z_1,\bar z_1)+\phi_4(z_1,\bar z_1)+\theta_3(z_1,\bar z_1)-\theta_4(z_1,\bar z_1))} e^{i (\phi_3(z_2,\bar z_2)+\phi_4(z_2,\bar z_2)+\theta_3(z_2,\bar z_2)-\theta_4(z_2,\bar z_2))}\big]/ 4 \,, \nn
&=\frac{1}{2|z/\alpha|^{(K_3+K_4+1/K_3+1/K_4)/2}} -\frac{\alpha^2 [ J_{\phi_3} \bar J_{\phi_3}+  J_{\theta_3} \bar J_{\theta_3}+  J_{\phi_4} \bar J_{\phi_4}+ J_{\theta_4} \bar J_{\theta_4}]_{(z_c,\bar z_c)}}{2|z/\alpha|^{(K_3+K_4+1/K_3+1/K_4)/2 -2}} +\cdots
\label{dc1},\\
\cos[\phi_3(z_1,\bar z_1)+\phi_4(z_1,\bar z_1)&-\theta_3(z_1,\bar z_1)+\theta_4(z_1,\bar z_1)] \cos[\phi_3(z_2,\bar z_2)+\phi_4(z_2,\bar z_2)-\theta_3(z_2,\bar z_2)+\theta_4(z_2,\bar z_2)] \nn 
&= \big[e^{i (\phi_3(z_1,\bar z)+\phi_4(z_1,\bar z_1)-\theta_3(z_1,\bar z_1)+\theta_4(z_1,\bar z_1))}e^{-i (\phi_3(z_2,\bar z_2)+\phi_4(z_2,\bar z_2)-\theta_3(z_2,\bar z_2)+\theta_4(z_2,\bar z_2))}\nn 
&~~~~~~+e^{-i (\phi_3(z_1,\bar z_1)+\phi_4(z_1,\bar z_1)-\theta_3(z_1,\bar z_1)+\theta_4(z_1,\bar z_1))} e^{i (\phi_3(z_2,\bar z_2)+\phi_4(z_2,\bar z_2)-\theta_3(z_2,\bar z_2)+\theta_4(z_2,\bar z_2))}\big]/ 4 \,, \nn
&=\frac{1}{2|z/\alpha|^{(K_3+K_4+1/K_3+1/K_4)/2}} -\frac{\alpha^2 [ J_{\phi_3} \bar J_{\phi_3}+  J_{\theta_3} \bar J_{\theta_3}+  J_{\phi_4} \bar J_{\phi_4}+ J_{\theta_4} \bar J_{\theta_4}]_{(z_c,\bar z_c)}}{2|z/\alpha|^{(K_3+K_4+1/K_3+1/K_4)/2 -2}} +\cdots\,.
\label{dc2}
\end{align}
Hence, the scaling dimension of the involved cosines in the $\tilde \Delta_c$ term is given by $(K_3+K_4+1/K_3+1/K_4)/4$, and thus,  the corresponding RG flow equations are written as $\frac{d \tilde\Delta_c}{dl}=[2-(K_3+K_4+1/K_3+1/K_4)/4]\tilde\Delta_c$. The sum of the second terms in Eqs. (\ref{dc1}),(\ref{dc2}) gives the renormalization of the LL parameters $K_3$ and $K_4$. We follow the procedure defined in Eq. (\ref{ed3}-\ref{ed5}) and compute the contributions to the $K_3$ flow from the term $J_{\phi_3} \bar J_{\phi_3}$ as $dK_3^{-1}=\frac{ \tilde\Delta_c^2}{2} dl$ and from the  term $J_{\theta_3} \bar J_{\theta_3}$  as $dK_3=\frac{ \tilde \Delta_c^2}{2} dl$.
Summing up these two contributions, we arrive at $ \frac{dK_3}{dl}=\frac{ (1-K_3^2) \tilde \Delta_c^2}{2} $. Similarly, we find the flow for the LL parameter $K_4$, $\frac{dK_4}{dl}=\frac{ (1-K_4^2) \tilde\Delta_c^2}{2}$.
In the same way, we calculate the OPE coefficient for the remaining terms in the Hamiltonian and get the RG flow equations displayed in Eq. (\ref{rg1}) of the main text.

\section{Derivation of RG equations in the microscopic source term approach}
\label{argt}

In this Appendix, we compute the RG flow equations starting from the tunneling Hamiltonian given by Eq. (\ref{HT}). It describes the tunneling between each of two NWs and the three-dimensional $s$-wave superconductor. For simplicity, we assume that the strength of electron-electron interactions is the  same in the two NWs such that $K_1=K_2$ and $K_3=K_4$. In the following section, we explicitly start from the partition function containing both NWs and SC degrees of freedom. By integrating out the SC part, we calculate the contribution to the pairing terms induced in the NWs by the tunneling terms.

\subsection{ Terms associated with tunneling  $t_{int}$ to interior branches of the spectrum }
\subsubsection{Contribution to the direct superconducting pairing $\Delta_\tau^{int}$ induced at interior branches of the spectrum}
\label{contri1}

\begin{figure}
        \centering
                \includegraphics[width=0.5 \textwidth]{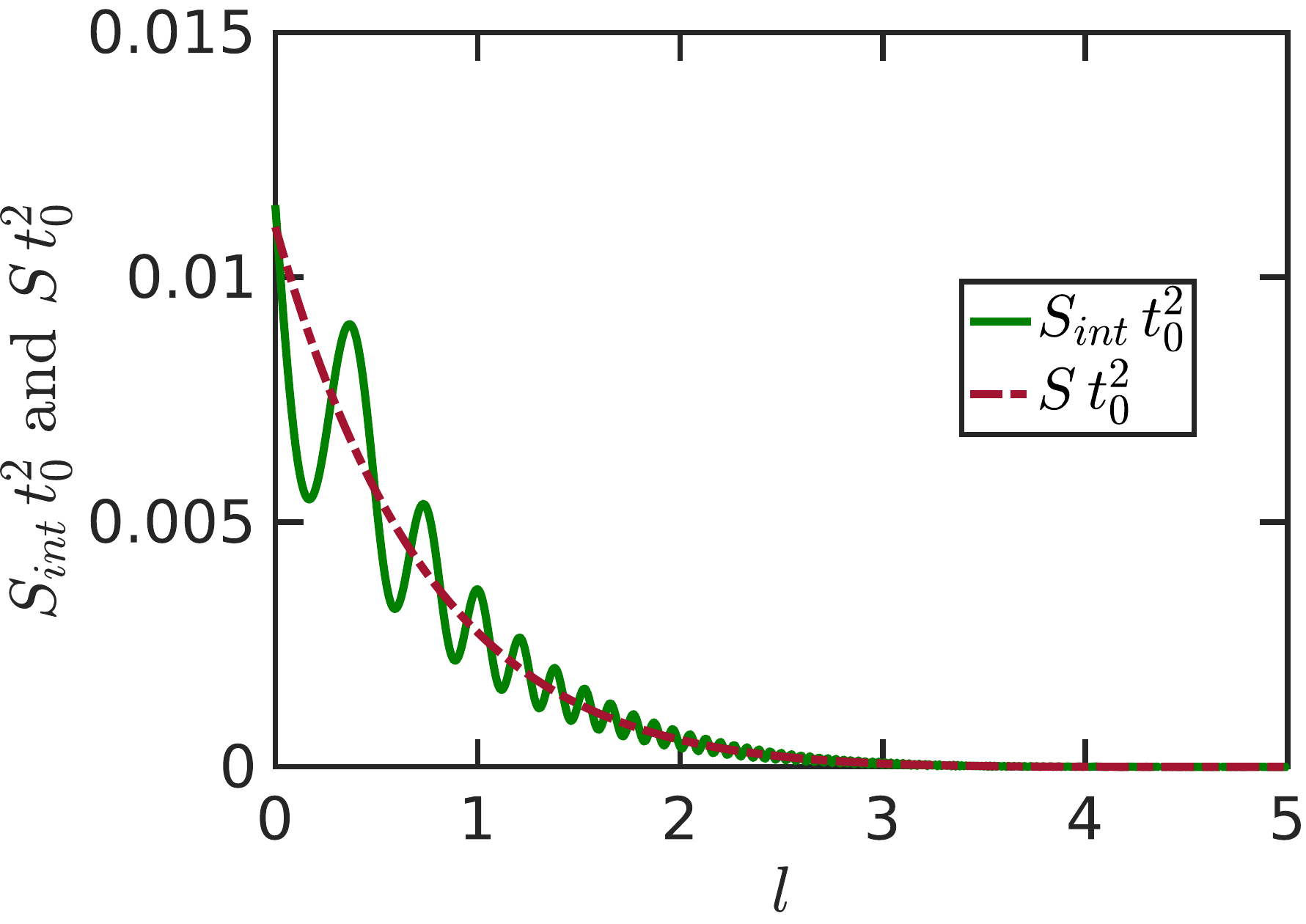} 
        \caption{ The comparison between the two source terms  $S_{int} \,t_0^2$ (green line) and  $S \,t_0^2$ (red dashed) represented as a function of the RG flow parameter $l$, which are calculated using Eq. (\ref{sexact}) and  Eq. (\ref{sapprox}), respectively. Disregarding small oscillations, the agreement between the two terms is fairly good.
        The parameter values are fixed to $t_0=\tilde t_{int}(0)=\tilde t_{ext}(0)=3.8 \times 10^{-5}$, $m_e$ is electron mass,
$\Delta=0.35\, \text{meV}$,
$u=10^4~ \text{m/s}$,
$v_{F,sc}= 10^6~ \text{m/s}$, $\alpha_0=1 \,\text{nm}$,  $d=15 \,\alpha_0$, and $\alpha_{sc}=1/k_{F,sc}=1 \buildrel _\circ \over {\mathrm{A}}$.}
\label{fig_S_ap}
\end{figure}

In the following subsection, we calculate the contribution from the tunneling Hamiltonian given by Eq. (\ref{HT}) to the flow equation of $\Delta_\tau^{int}$. In the partition function $Z$, we expand the action up to second order in the tunneling term,  which  results in a first-order contribution to the proximity-induced superconducting pairing of the type 
$\sum_{\tau}  \frac{\tilde\Delta^{ int}_{\tau}u}{\alpha}
\int dx\,  ( R_{\tau\bar1} L_{\tau 1} +L_{\tau 1}^\dagger R_{\tau\bar1}^\dagger) $. Here, $\tilde\Delta^{ int}$ is a dimensionless  coupling constant, like in Eq. (\ref{HSC1}), which is initially zero but then assumes a finite value during the RG procedure as shown below. Without lost of generality, we focus on the first term in the first NW, {\it i.e.} on $R_{1\bar1} L_{1 1}$, which in terms of the bosonic operator has the form   
$\frac{\tilde\Delta^{ int}_{1}u}{2 \pi \alpha^2} \int dx \, e^{2 i \phi_3(x,t)} $.
The partition function\cite{TG,Cardy} (dimensionless) can be written as 
\begin{align}
Z= Z_0 \,\Big\langle\, 1-  \frac{\tilde\Delta^{ int}_{1}u}{2 \, \pi \, \alpha^2}\int  dx \,dt\, [e^{2 i \phi_3(x,t)}+\text{H.c.}] + I+I^\dagger + \cdots  \Big\rangle_0 \,,
\label{Zin}\end{align} 
where $Z_0$ is the partition function corresponding to the fixed point Hamiltonian $H_0$ written in Eq. (\ref{aH0}),  and $\langle \,\cdots\,\rangle_0$ denotes the expectation value over the NWs with respect to  $H_0$.
 The second order term $I$ in the partition function, coming from the tunneling Hamiltonian given by Eq. (\ref{HT}), is rewritten as 
\begin{align}
I=\frac{1}{2}&  \int \,dx_1\, dt_1\, dx_2\, dt_2\, d{\bf r}\, d{\bf r}'\,  u^2\,\Big(\frac{\xi^2 \, L}{\alpha^3}\Big)~ \tilde t^2_{int} \,T[R_{1\bar1}(x_1,t_1) L_{11}(x_2,t_2)]\, \langle 
T[\Psi^\dagger_{\downarrow}({\bf r},t_1)\Psi^\dagger_{\uparrow}({\bf r}',t_2)]\rangle \nn  
&\hspace{40pt}\times \delta(r_x-x_1) \delta(r_y) \delta(r_z) \delta(r'_x-x_2) \delta(r'_y) \delta(r'_z) ,
\end{align}
where $R_{\tau \sigma}(x)$ and $L_{\tau \sigma}(x)$ are slowly varying right and left moving fields with spin $\sigma/2$ in the $\tau $-NW at position $x$
and $\langle \,\cdots\,\rangle$ denotes the  equilibrium expectation value over the degrees of freedom of the SC.
 We write $R_{1\bar1}(x_1,t_1)$ and $L_{11}(x_2,t_2)$ in bosonic language as $e^{ i [\phi_3(x_1,t_1)-\theta_3(x_1,t_1)]}/\sqrt{2 \, \pi\,\alpha}$ and $e^{ i [\phi_3(x_2,t_2)+\theta_3(x_2,t_2)]}/\sqrt{2 \, \pi\,\alpha}$. Thus, $I$ can be written as
\begin{align}
I=&\frac{1}{4 \, \pi \,} \int \,dx_1\, dt_1 \,dx_2\, dt_2 \,d{\bf r} \,d{\bf r}'\, u^2\Big(\frac{\xi^2 \, L}{\alpha^4}\Big)~\tilde t^2_{int}T[ e^{ i [\phi_3(x_1,t_1)-\theta_3(x_1,t_1)]} e^{ i [\phi_3(x_2,t_2)+\theta_3(x_2,t_2)]}]  \nn
&\hspace{40pt}  \times \langle T[\Psi^\dagger_{\downarrow}({\bf r},t_1)\Psi^\dagger_{\uparrow}({\bf r}',t_2)]\rangle \delta(r_x-x_1) \delta(r_y) \delta(r_z) \delta(r'_x-x_2) \delta(r'_y) \delta(r'_z) .
\label{I3}
\end{align}
The anomalous Green function for our $s$-wave superconductor is given by $
\langle T[\Psi_{\sigma_1}({\bf r},t_1)\Psi_{\sigma_2}({\bf r}',t_2)]\rangle=\sigma_1 \delta_{\sigma_1,-\sigma_2} F({\bf r},{\bf r}',t_1,t_2)$\cite{Recher}, where the $F$-function can be calculated from its inverse Fourier transform function
\begin{align}
F({\bf r},{\bf r}',t_1,t_2)=\int \int \  \frac{d\omega\, d{\bf k}}{(2\pi)^4} e^{i( {\bf r}- {\bf r}')\cdot{\bf k}}e^{i \omega (t_1-t_2)} \frac{\Delta}{\omega^2+E_k^2+\Delta^2}.
\end{align}
Therefore, the anomalous Green function becomes
\begin{align}
 \langle T[\Psi^\dagger_\downarrow({\bf r},t_1)\ \, \Psi^\dagger_{\uparrow }({\bf r}',t_2)]  \rangle =-
&\int \frac{d\omega \, d{\bf k} \, }  {(2 \, \pi)^4 }
\frac{\Delta e^{i({\bf r}- {\bf r}')\cdot{\bf k}+i \omega (t_1-t_2) }} {\omega^2 +\Delta^2+ [(k^2-k_{F,sc}^2 )/2m_e]^2   } \, \nn
 & = - \int    d \omega  \,
  \frac{m_e\,\Delta\, e^{-\frac{ |r-r'|   \sqrt{\Delta^2 +   \omega^2  } }{v_{F,sc}}+ i \omega (t_1-t_2)}    \, \sin (k_{F,sc}\, |r-r'|)
 }
 { (2\, \pi)^2\,  |r-r'|\, \sqrt{\Delta^2 +   \omega^2  }  }\,,
\end{align}
where $m_e$, $v_{F,sc}$, and $k_{F,sc}$ are the electron mass, Fermi velocity, and Fermi wavevector of the bulk SC, respectively. Hence, Eq. (\ref{I3}) can be rewritten as

\begin{align}
I=& -\frac{1}{4 \, \pi \,} u^2\Big(\frac{\xi^2 \, L}{\alpha^4}\Big)~\tilde t^2_{int} \int   \,dx_1\, dt_1\, dx_2\, dt_2\,  T[e^{ i [\phi_3(x_1,t_1)-\theta_3(x_1,t_1)]} \,e^{ i [\phi_3(x_2,t_2)+\theta_3(x_2,t_2)]}]  \nn
& \hspace*{2cm}\times \int    d \omega 
  \frac{m_e\,\Delta\, e^{-\frac{ |x_1-x_2|   \sqrt{\Delta^2 +   \omega^2  } }{v_{F,sc}}+ i \omega |t_1-t_2|}   \sin (k_{F,sc}\, |x_1-x_2|)
 } { (2\, \pi)^2\,  |x_1-x_2|\, \sqrt{\Delta^2 +   \omega^2  }  }
    . \label{A1}
\end{align}
For low energy modes,  we can approximate $e^{-\frac{ | x_1 - x_2|\,\sqrt{\Delta^2 + \omega^2 } } {v_{F,sc} }} \approx e^{-\frac{ |x_1-x_2|\,\Delta}  {v_{F,sc} }}$, thus 
\begin{align}
I= & 
-\frac{m_e \,\Delta\, u^2\, \tilde t^2_{int} \xi^2 \, L}{16 \,\pi^3\, \alpha^4} \int \,dx_1 \,dt_1\, dx_2\, dt_2\, T[e^{ i [\phi_3(x_1,t_1)-\theta_3(x_1,t_1)]}\, e^{ i [\phi_3(x_2,t_2)+\theta_3(x_2,t_2)]}]  \frac{\sin (k_{F,sc}\, |x_1-x_2|)~ e^{-\frac{ |x_1-x_2|\,\Delta}  {v_{F,sc} }}}{ |x_1-x_2|} \nn
&\hspace*{3cm}\times \int d \omega 
  \frac{e^{ i \omega  |t_1-t_2|}} { \sqrt{\Delta^2 +   \omega^2  }  }  ,\\
  = & 
-\frac{m_e\, \Delta \,u^2\, \tilde t^2_{int}\,\xi^2 \, L}{16 \,\pi^3\, \alpha^4} \int  \,dx_1 \,dt_1 \,dx_2\, dt_2\, T[ e^{ i [\phi_3(x_1,t_1)-\theta_3(x_1,t_1)]} \,e^{ i [\phi_3(x_2,t_2)+\theta_3(x_2,t_2)]}]  \frac{\sin (k_{F,sc}\, |x_1-x_2|)~ e^{-\frac{ |x_1-x_2|\,\Delta}  {v_{F,sc} }}}{ |x_1-x_2|}\nn
& \hspace*{3cm} \times 2\, K_0( \, |  t_1-t_2|   \, \Delta )  ,
\end{align}
where $K_0( p)=\int_0^\infty dx \frac{\text{cos}(px)}{\sqrt{x^2+1}}$ is the modified Bessel function of the second kind. Furthermore, $\Delta$ is assumed to be $\omega$-independent. At the next step, we change to the center-of-mass coordinates $(X,T,x,t)$, where $X=(x_1+x_2)/2$, $T=(t_1+t_2)/2$, $x=x_1-x_2$, and $t=t_1-t_2$. We also define complex coordinates $z_{1/2}=-i \,x_{1/2} + u \,t_{1/2}$, $z_c= (z_1+z_2)/2$, and $z=z_1-z_2$. As a result, we get
\begin{align}
I=-\frac{m_e \Delta u^2 \,\tilde t^2_{int} \xi^2  L}{8 \,\pi^3\, \alpha^4} \int  \,dX \,dT\,dx\, dt \, T[e^{ i [\phi_3(x_1,t_1)-\theta_3(x_1,t_1)]}\, e^{ i [\phi_3(x_2,t_2)+\theta_3(x_2,t_2)]}] \frac{\sin (k_{F,sc}\, |x|)~ e^{-\frac{ |x|\,\Delta}  {v_{F,sc} }}}{ |x|} K_0(  |  t|    \Delta )     . \label{Iint}
\end{align} 
To begin with the RG analysis, we again change the cutoff from $\alpha \rightarrow \alpha+d\alpha$ (where $d\alpha=  \alpha dl$). The integral over $(x,t)$ is converted to an integral over polar coordinates $(r,\theta')$. Next, we split the integral over $r$ into two parts $\int_{r>(\alpha+d\alpha)} = \int_{r>\alpha}-\int_{\alpha}^{\alpha+d \alpha}$. The first one gives the original integral in the partition function and we only need to compute the integral $I$ within $\alpha < r < \alpha + d\alpha$ . Again, we make use of following OPEs\cite{Cardy} written in terms of the complex coordinates $(z_1,\bar z_1)$ and $(z_2,\bar z_2)$ with $z_c=(z_1+z_2)/2$ and $z=z_1-z_2$, 
\begin{align}
 T \left[e^{ i [\phi_3(z_1,\bar z_1)-\theta_3(z_1,\bar z_1)]} \, e^{ i [\phi_3(z_2,\bar z_2)+\theta_3(z_2,\bar z_2)]} \right]~~
   \overset{(z_1,\bar z_1) \rightarrow (z_2,\bar z_2)}{=}~~   \frac{1}{ |z/\alpha|\, ^{s_r+s_l-s}} 
    e^{2 i \phi_3 (z_c,\bar z_c)}+ \cdots,
 \end{align}
where $s_r=s_l=(K_3+1/K_3)/4$ and $s=K_3$ are the scaling dimensions of $e^{ i [\phi_3(z,\bar z)-\theta_3(z,\bar z)]},e^{ i [\phi_3(z,\bar z)+\theta_3(z,\bar z)]}$, and $e^{2 i \phi_3 (z,\bar z)}$, respectively. As a result, Eq. (\ref{Iint}) can be rewritten  as
\begin{align}
I^{d\alpha}=-\frac{m_e \,\Delta\, u^2 \,\tilde t^2_{int}\, \xi^2 \, L}{8 \,\pi^3\, \alpha^4} \int  \,dX \,dT\,dx\, dt \frac{1}{|z/\alpha|^{s_r+s_l-s}} e^{2 i \phi_3(X,T)}\frac{\sin (k_{F,sc}\, |x|)~ e^{-\frac{ |x_1-x_2|\,\Delta}  {v_{F,sc} }}}{ |x|} K_0( \, |  t|   \, \Delta )      . 
\label{Iint1}\end{align}
For simplicity, we define $I_1$ such that
\begin{align}
I^{d\alpha}=&\int dX \,dT e^{2 i \phi_3(X,T)} I_1\,, \label{Iint2}\\
I_1=&-\frac{m_e \,\Delta\, u^2\, \tilde t^2_{int}\,\xi^2 \, L}{8 \,\pi^3\, \alpha^4} \int   \, dx \,dt   \frac{1}{|z/\alpha|^{s_r+s_l-s}} \frac{\sin (k_{F,sc}\, |x|)~ e^{-\frac{ |x_1-x_2|\,\Delta}  {v_{F,sc} }}}{ |x|} K_0( \, |  t|   \, \Delta )    \nn 
=&-\frac{m_e\, \Delta\, u\, \tilde t^2_{int}\,\xi^2 \, L}{ 2 \,\pi^3\, \alpha^4}
\int  \int_0^{\pi/2} \frac{r \,dr \,  d\theta' \,}{|r/\alpha|^{s_r+s_l-s}}
\sin\left(  k_{F,sc}\, | r~\text{cos}(\theta') | \right)
 \frac{e^{-\frac{|r~\text{cos}(\theta')|  \, \Delta  }{v_{F,sc} } }}{| r~ \text{cos}(\theta')|}\, 
 K_0\left( \frac{ |r~ \text{sin}(\theta')| \Delta   } {u} \right).
\end{align}
We only need to compute the integral $I_1$ within $\alpha < r < \alpha + d\alpha$ as described above.  Using $\int_\alpha^{\alpha+ d\alpha} dr f(r) = d\alpha \,f(\alpha)$, we get the simplified expression as follows
\begin{align}
I_1 = \Big(\frac{u \,d\alpha}{\alpha^{3}}\Big) \frac{m_e \,\Delta \,\tilde t^2_{int} \,\xi^2 \, L}{  2 \, \pi^3 \, \alpha} 
 \int_0^{\pi/2}   d\theta' \,
\sin\left(  k_{F,sc}\, | \alpha~\text{cos}(\theta') | \right)
 \frac{e^{-\frac{|\alpha~\text{cos}(\theta')|  \, \Delta  }{v_{F,sc} } }}{| \text{cos}(\theta')|}\, 
 K_0\left( \frac{ |\alpha~ \text{sin}(\theta')| \Delta   } {u} \right).
\end{align}
Therefore, Eq. (\ref{Iint2}) takes the form
\begin{align}
I^{d\alpha} =&\left[ \frac{ m_e \Delta \tilde t^2_{int} \xi^2  L}{ \pi^2 \alpha} 
 \int_0^{\pi/2}   d\theta' \,
\sin\left(  k_{F,sc} | \alpha\cos(\theta') | \right)
 \frac{e^{-\frac{|\alpha \cos(\theta')|  \Delta  }{v_{F,sc} } }}{| \cos(\theta')|}
 K_0\left( \frac{ |\alpha \sin(\theta')| \Delta   } {u} \right)\right] \int \frac{u}{2 \, \pi \,\alpha^{2}} dX \,dT e^{2 i \phi_3(X,T)}  \frac{d\alpha}{ \alpha}.
 \label{If}
\end{align}
This term allows us to find the contribution to the direct superconducting pairing $\tilde{\Delta}_\tau^{int}$ in the first order, see Eq.~(\ref{Zin}). Since $dl= \frac{d\alpha}{ \alpha} $ and $\xi=v_{F,sc}/ \Delta$, this leads us to 
\begin{align}
\frac{d\tilde{\Delta}_\tau^{int}}{dl}&=S_{int} \, \tilde t^2_{int} \,,\\
S_{int}&=\frac{ m_e \,v_{F,sc}^2 \, L}{ \pi^2 \, \Delta \, \alpha} 
 \int_0^{\pi/2}   d\theta' \,
\sin\left(  k_{F,sc}\, | \alpha~\text{cos}(\theta') | \right)
 \frac{e^{-\frac{|\alpha~\text{cos}(\theta')|  \, \Delta  }{v_{F,sc} } }}{| \text{cos}(\theta')|}\, 
 K_0\left( \frac{ |\alpha~ \text{sin}(\theta')| \Delta   } {u} \right).
 \label{sexact}
\end{align}
Next, we estimate the integral over the polar angle by noting that the main contribution comes from angles close to $\pi/2$. We consider $\int_0^{\pi/2}=\int_{\theta'_c}^{\pi/2}+ \, \text{small contribution}$, where $\theta'_c= \text{cos}^{-1}(\frac{\pi}{2\,k_{F,sc} \alpha})=\pi/2 - \text{sin}^{-1}(\frac{\pi}{2\,k_{F,sc}\alpha})=\pi/2 \,[1- (1/k_{F,sc}\alpha)]$. We use $\int_{\theta'_1}^{\theta'_1-d\theta'} f(\theta')= -f(\theta'_1) \, d\theta'$ and approximate the integral in $S_{int}$ as
 \begin{align}
 S&=\frac{ m_e \,v_{F,sc}^2 \, L}{ \pi^2 \, \Delta \, \alpha} 
 \int_{\pi/2 [1- (1/k_{F,sc}\alpha)]}^{\pi/2}   d\theta' \,
\sin\left(  k_{F,sc}\, | \alpha~\text{cos}(\theta') | \right)
 \frac{e^{-\frac{|\alpha~\text{cos}(\theta')|  \, \Delta  }{v_{F,sc} } }}{| \text{cos}(\theta')|}\, 
 K_0\left( \frac{ |\alpha~ \text{sin}(\theta')| \Delta   } {u} \right) \nn
 &=\frac{ m_e \,v_{F,sc}^2  \, L}{  2\, \pi\, \Delta \, \alpha} \,
 K_0\left( \frac{ \alpha\, \Delta   } {u} \right).
 \label{sapprox}
 \end{align}
At the last step, we restore $\hbar$-factors and arrive at $S=\frac{ m_e \,v_{F,sc}^2  \, L}{  2\, \pi\, \Delta \, \alpha} \,
 K_0\left( \frac{ \alpha\, \Delta   } {\hbar \,u} \right)$. In Fig. \ref{fig_S_ap}, we demonstrate that the approximate value $S\, t_0^2$ of the contribution coming from the tunneling term matches nicely with its exact value $S_{int} t_0^2$ found numerically. Thus, for simplicity, we can use $S$ instead of $S_{int}$ for the numerical evaluation of the RG flow equations.
 
\subsubsection{Contribution to the crossed Andreev superconducting pairing $\Delta_c$}
\label{contri2}

In this subsection, we calculate the contribution from the tunneling Hamiltonian given by Eq. (\ref{HT}) to the flow equation of $\Delta_c$. In the partition function $Z$, we expand again the action up to second order in the tunneling term,  which  results in a first-order contribution to the proximity-induced superconducting pairing 
\begin{align}
Z= Z_0 \,\Big\langle \, 1-  \frac{\tilde\Delta_{c}u}{2 \, \pi \,\alpha^2}\int dx \,dt \, (e^{2 i [\phi_3(x,t)-\theta_3 (x,t)+\phi_4 (x,t)+\theta_4 (x,t)]}+ \text{H.c.}) +  I_c +I_c^\dagger  \cdots   \,\Big\rangle_0.
\label{Zc}\end{align}
Similar to the previous subsection, we introduce and compute $I_c$ as was done before for $I$ in Eq. (\ref{A1}),
\begin{align}
I_c= & 
-\frac{1}{4\,\pi} \int dx_1\,dt_1\, dx_2 \,dt_2\, d{\bf r}\, d{\bf r}'\, u^2\Big(\frac{\xi^2 \, L}{\alpha^4}\Big)~\tilde t^2_{int}T[e^{ i [\phi_3(x_1,t_1)-\theta_3(x_1,t_1)]} \,e^{ i [\phi_4(x_2,t_2)+\theta_4(x_2,t_2)]}] \nn 
 &\times \int    d \omega 
  \frac{m_e\,\Delta\, e^{-\frac{ |r-r'|   \sqrt{\Delta^2 +   \omega^2  } }{v_{F,sc}}+ i \omega |t_1-t_2|}   \sin (k_{F,sc}\, |r-r'|)
 } { (2\, \pi)^2\,  |r-r'|\, \sqrt{\Delta^2 +   \omega^2  }  } \delta(r_x-x_1) \delta(r_y) \delta(r_z) \delta(r'_x-x_2) \delta(r'_y-d) \delta(r'_z) \nn
=&-\frac{1}{4\,\pi} \int dx_1 \,dt_1 \,dx_2\, dt_2\, u^2\Big(\frac{\xi^2 \, L}{\alpha^4}\Big)~\tilde t^2_{int}T[e^{ i [\phi_3(x_1,t_1)-\theta_3(x_1,t_1)]} \, e^{ i [\phi_4(x_2,t_2)+\theta_4(x_2,t_2)]}]  \frac{\sin (k_{F,sc}\, \sqrt{|x_1-x_2|^2+d^2})}{(2\, \pi)^2\,  \sqrt{|x_1-x_2|^2+d^2}\,} \nn
&\times  \int    d \omega 
  \frac{m_e\,\Delta\, e^{-\frac{ \sqrt{|x_1-x_2|^2+d^2}   \sqrt{\Delta^2 +   \omega^2  } }{v_{F,sc}}+ i \omega |t_1-t_2|}   
 } {  \sqrt{\Delta^2 +   \omega^2  }  } . 
\end{align}
For low energy modes,  we again approximate $e^{-\frac{ \sqrt{|x_1-x_2|^2+d^2}\,\sqrt{\Delta^2 + \omega^2 } } {v_{F,sc} }}$ by $e^{-\frac{ \sqrt{|x_1-x_2|^2+d^2}\,\Delta}  {v_{F,sc} }}$, resulting in 
\begin{align}
I_c= & 
-\frac{m_e \,\Delta\, u^2\, \tilde t^2_{int}\, \xi^2 \, L}{16\, \pi^3 \,\alpha^4} \int\,dx_1 \,dt_1 \,dx_2 \,dt_2 ~T [ e^{ i [\phi_3(x_1,t_1)-\theta_3(x_1,t_1)]}\,e^{ i [\phi_4(x_2,t_2)+\theta_4(x_2,t_2)]} ]  \nn 
& \hspace*{2.5cm}\times \frac{\sin (k_{F,sc}\, \sqrt{|x_1-x_2|^2+d^2})~ e^{-\frac{ \sqrt{|x_1-x_2|^2+d^2}\Delta}  {v_{F,sc} }}}{ \sqrt{|x_1-x_2|^2+d^2}}  \int  d \omega 
  \frac{e^{ i \omega  |t_1-t_2|}} { \sqrt{\Delta^2 +   \omega^2  }  }    \nn
  = & 
-\frac{m_e \,\Delta\, u^2\, \tilde t^2_{int}\, \xi^2 \, L}{16 \,\pi^3 \,\alpha^4} \int \,dx_1 \,dt_1 \,dx_2\, dt_2\, T[e^{ i [\phi_3(x_1,t_1)-\theta_3(x_1,t_1)]} \,e^{ i [\phi_4(x_2,t_2)+\theta_4(x_2,t_2)]}] \nn & \hspace*{2.5cm}\times \frac{\sin (k_{F,sc}\, \sqrt{|x_1-x_2|^2+d^2})~ e^{-\frac{ \sqrt{|x_1-x_2|^2+d^2}\Delta}  {v_{F,sc} }}}{ \sqrt{|x_1-x_2|^2+d^2}} 2\, K_0( \, |  t_1-t_2|   \, \Delta )    ,
\label{D21}
\end{align}
where $K_0( p)=\int_0^\infty dx \frac{\text{cos}(px)}{\sqrt{x^2+1}}$ is the modified Bessel function of the second kind.
We assume again $\Delta$ to be $\omega$-independent, which is a good approximation for our low-energy theory. We change the coordinates to center-of-mass coordinates $X=(x_1+x_2)/2$, $T=(t_1+t_2)/2$, $x=x_1-x_2$, and $t=t_1-t_2$. We also introduce $(z_1+z_2)/2=z_c$ and $z=(z_1-z_2)$, where $(z_1,\bar z_1)$ and $(z_2,\bar z_2)$ are the complex coordinates as defined in previous Sec. \ref{contri1} . We again change the cutoff from $\alpha$ to $\alpha+ d\alpha$ for the RG analysis and switch to for $(x,t)$ to the polar coordinates $(r,\theta')$, such that we need to calculate the integral only for $\alpha<r<\alpha+d\alpha$. We make use of the following OPE:
\begin{align}
T[e^{ i [\phi_3(z_1,\bar z_1)-\theta_3(z_1,\bar z_1)]} \,e^{ i [\phi_4(z_2,\bar z_2)+\theta_4(z_2,,\bar z_2)]}] &\overset{(z_1,,\bar z_1)\rightarrow (z_2,,\bar z_2)}{=}~~   \frac{1}{ |z/\alpha|^{c_r+c_l-c}}   \nn
   & \hspace*{1cm} \times e^{ i \phi_3 (z_c,\bar z_c)-\theta_3(z_c,\bar z_c)+\phi_4 (z_c,\bar z_c)+\theta_4 (z_c,\bar z_c)}+ \cdots.
\end{align}
Here the scaling dimensions of $e^{ i \phi_3 (z,\bar z)-\theta_3 (z,\bar z)}$,  $e^{ i \phi_4 (z,\bar z)+\theta_4 (z,\bar z)}$,  and $e^{[ i \phi_3 (z,\bar z)-\theta_3 (z,\bar z)+\phi_4 (z,\bar z)+\theta_4 (z,\bar z)]}$ are $c_r=(K_3+1/K_3)/4$, $c_l=(K_4+1/K_4)/4$, and $c=(K_3+1/K_3+K_4+1/K_4)/4$, respectively.   This yields the following correction coming from Eq. (\ref{D21}) to the partition function,
\begin{align}
I_c^{d\alpha}&=-\frac{m_e \,\Delta \,u^2\, \tilde t^2_{int} \,\xi^2 \, L}{8 \, \pi^3\, \alpha^4} \int \,dX \,dT \,dx \, dt  \frac{1}{|z/\alpha|^{c_r+c_l-c}} e^{[ i \phi_3 (X,T)-\theta_3 (X,T)+\phi_4 (X,T)+\theta_4 (X,T)]} \nn
&\hspace{40pt} \times \frac{\sin (k_{F,sc}\, \sqrt{|x|^2+d^2})~ e^{-\frac{ \sqrt{|x|^2+d^2}\Delta}  {v_{F,sc} }}}{ \sqrt{|x|^2+d^2}}  K_0( \, |  t|   \, \Delta )  .
\end{align}
As a result, we rewrite $I_c$ as
\begin{align}
I_c^{d\alpha}&=
\int dX \,dT e^{[ i \phi_3 (X,T)-\theta_3 (X,T)+\phi_4 (X,T)+\theta_4 (X,T)]} I_1^{d\alpha}\,,\label{I}\\
I_1^{d\alpha}&=-\frac{m_e \,\Delta\, u^2 \,\tilde t^2_{int}\xi^2 \,L}{8 \, \pi^3\, \alpha^4} \int dx \, dt\, \frac{1}{|z/\alpha|^{c_r+c_l-c}} \frac{\sin (k_{F,sc}\, \sqrt{|x|^2+d^2})~ e^{-\frac{ \sqrt{|x|^2+d^2}\Delta}  {v_{F,sc} }}}{\sqrt{|x|^2+d^2}} K_0( \, |  t|   \, \Delta )       \\
 &=-\frac{m_e \,\Delta \,u \,\tilde t^2_{int}\xi^2 \, L}{ 2\, \pi^3\, \alpha^4}
\int \int_0^{\pi/2} \frac{r\, dr \,  d\theta' \,}{|r/\alpha|^{c_r+c_l-c}}
\sin\left(  k_{F,sc}\, \sqrt{ r^2~\text{cos}^2(\theta')+d^2} \right)
 \frac{e^{-\frac{\sqrt{ r^2~\text{cos}^2(\theta')+d^2} \Delta  }{v_{F,sc} } }}{\sqrt{ r^2~\text{cos}^2(\theta')+d^2}}\, 
 K_0\left( \frac{ |r~ \text{sin}(\theta')| \Delta   } {u} \right). \nonumber
\end{align}
We compute the integral $I_1^{d\alpha}$ within $\alpha <  r < \alpha + d\alpha$  (where $d\alpha= \alpha ~dl$). Using $\int_\alpha^{\alpha+ d\alpha} dr f(r) = d\alpha f(\alpha)$ and considering the fact that $d\gg \alpha$ (we keep only the most singular terms in $\alpha$ small before we scale $\alpha$ up), we get
\begin{align}
I_1^{d\alpha} = \Big(\frac{u \, d\alpha}{\alpha^{3}} \Big) \frac{m_e \,\Delta\, \tilde t^2_{int} \, \xi^2 \, L}{ 2\,\pi^2 }
 \int_0^{\pi/2}   d\theta' \,
\sin\left(  k_{F,sc}\, d \right)
 \frac{e^{-\frac{d  }{\xi } }}{d}\, 
 K_0\left( \frac{ |\alpha~ \text{sin}(\theta')| \Delta   } {u} \right),
\end{align}
which leads us to  $I_c^{d\alpha}$ in the form
\begin{align}
I_c^{d\alpha} =& \frac{u}{2\, \pi\,\alpha^{2}} \int dX \,dT \,e^{[ i \phi_3 (X,T)-\theta'_3 (X,T)+\phi_4 (X,T)+\theta_4 (X,T)]} \nn
 &\hspace*{1cm}\times \left[\frac{m_e \,\Delta \,\tilde t^2_{int} \, \xi^2 \,L\, e^{-\frac{d }{\xi} }\sin\left(  k_{F,sc}\, d \right)}{\pi^2\, d}\,
 \int_0^{\pi/2}   d\theta' \, 
 K_0\left( \frac{ |\alpha~ \text{sin}(\theta')| \Delta   } {u} \right)\right] \frac{d\alpha}{ \, \alpha} .
\end{align}
Again, $I_c^{d\alpha}$ contributes to the crossed Andreev superconducting pairing $\tilde\Delta_c$ in first order.
With $dl= \frac{d\alpha}{ \alpha} $ $\Rightarrow  \alpha=\alpha_0 \,e^{l}$, and the coherence length given by $\xi=\frac{ v_{F,sc}}{ \Delta}$, we conclude with the following relation for the contribution to the flow in $\tilde\Delta_c$:
\begin{align}
 \frac{d\tilde\Delta_c}{dl} =\frac{m_e \, \tilde t^2_{int} \,v_{F,sc}^2\, L\, | \sin\left(  k_{F,sc}\, d \right)|\,
 e^{-\frac{d}{\xi} }\, }{ \pi^2 \,d\, \Delta }
 \int_0^{\pi/2}   d\theta' \,
 K_0\left( \frac{ |\alpha~ \text{sin}(\theta')| \Delta   } {u} \right)=S_c\, \tilde t^2_{int}.
\end{align}
After putting back the $\hbar$-factors, the expression for $S_c$ takes the form,
\begin{align}
S_c= \frac{m_e \,v_{F,sc}^2\,L\, | \sin\left(  k_{F,sc}\, d \right)|\,
 e^{-\frac{d }{\xi } }\, }{ \pi^2 \,d\, \Delta }
 \int_0^{\pi/2}   d\theta' \,
 K_0\left( \frac{ |\alpha~ \text{sin}(\theta')| \Delta   } {\hbar \,u} \right).
\end{align}

\subsection{Contribution to the direct superconducting pairing $\Delta_\tau^{ext}$ induced at exterior branches of the spectrum}

In this subsection, we calculate the contribution from the tunneling Hamiltonian given by Eq. (\ref{HT}) to the flow equation of $\Delta_\tau^{ext}$. In the partition function $Z$, we expand the action up to second order in the tunneling term,  which  results in a first-order contribution to the proximity-induced superconducting pairing of the type 
$\sum_{\tau}  \frac{\tilde\Delta^{ ext}_{\tau}u}{\alpha}
\int dx\,  ( R_{\tau 1}^\dagger L_{\tau\bar1}^\dagger + L_{\tau\bar1} R_{\tau 1}) $ [$\tilde\Delta^{ ext}$ 
 is a dimensionless  coupling constant, see Eq. (\ref{HSC2})]. Without loss of generality, we focus on the first term in the first NW, {\it i.e.} on $L_{1 \bar1} R_{1 1}$.
We again start from the partition function given by
\begin{align}
Z= Z_0 \,\Big\langle 1-  \frac{ \tilde\Delta^{ ext}_{1}u}{\pi \,\alpha^2}\int  \cos\big[ 2\phi_1(x,t)\big] \, dx \,dt+ I+ I^{\dagger}  + \cdots  \Big\rangle_0,
\label{Zex}\end{align}
where $I^{\dagger}$ is the Hermitian conjugate of the second order contribution $I$ written as
\begin{align}
I=\frac{1}{2}  \int \,dx_1\, dt_1\, dx_2\, dt_2\, d{\bf r}\, d{\bf r}' & u^2\,\Big(\frac{\xi^2 \, L}{\alpha^3}\Big)~ \tilde t^2_{int} \,T[L_{1\bar1}(x_1,t_1) R_{11}(x_2,t_2)]\, \langle 
T[\Psi^\dagger_{\downarrow}({\bf r},t_1)\Psi^\dagger_{\uparrow}({\bf r}',t_2)]\rangle \nn  
&\times \delta(r_x-x_1) \delta(r_y) \delta(r_z) \delta(r'_x-x_2) \delta(r'_y) \delta(r'_z).
\end{align}
We follow the same procedure as described above in previous subsections. Then, $I^{d\alpha}$ similarly to Eq. (\ref{Iint1}) is given by
\begin{align}
I^{d\alpha}=-\frac{m_e \,\Delta \,u^2 \,\tilde t^2_{ext}\,\xi^2 \, L}{8 \, \pi^3\, \alpha^4}  \int \,dX \,dT \,dx\, dt \frac{1}{|z/\alpha|^{s_r+s_l-s}} e^{2i\phi_1(X,T)} \frac{\sin (k_{F,sc}\, |x|)~ e^{i k_{F\tau}  x}~ e^{-\frac{ |x|\,\Delta}  {v_{F,sc} }}}{ |x|} K_0( \, |  t|   \, \Delta )     .
\label{Iext}
\end{align}
The only difference between Eq. (\ref{Iint1}) and Eq. (\ref{Iext}) is an extra factor of $e^{i k_{F\tau} x}$ and $\phi_3 \rightarrow \phi_1$.
We add both ($I^{d\alpha}$ and $(I^{d\alpha})^\dagger$ contributions to calculate the renormalization of $\tilde\Delta^{ ext}_{1}$, and get
\begin{align}
I^{d\alpha}+(I^{d\alpha})^\dagger=&\frac{m_e  \Delta  \tilde t^2_{ext} \xi^2  L}{ 8 \, \pi^3\,\alpha^4} \int dX \,dT\,dx \,dt  \frac{1}{|z/\alpha|^{s_r+s_l-s}} \frac{2  \sin (k_{F,sc}\, |x|)  \cos(k_{F\tau}\, x)~ e^{-\frac{ |x|\,\Delta}  {v_{F,sc} }}}{ |x|} K_0( \, |  t|   \, \Delta )     \cos[2\phi_1(X,T)]  .
\label{E1}
\end{align}
We follow the same procedure as used in Eqs. (\ref{Iint2}-\ref{If}) and calculate the integral in the range  $\alpha<r<\alpha+d \alpha$, as a result, Eq. (\ref{E1}) takes the following form:
\begin{align}
I^{d \alpha}+I^{d \alpha ^\dagger}=&\Big(\frac{u \,d\alpha}{\alpha^{3}}\Big) \frac{ m_e \,\Delta \,\tilde t^2_{ext} \,\xi^2 \, L}{ \pi^3 \, \alpha} 
 \int_0^{\pi/2}   d\theta' \,
\sin (k_{F,sc}\, |x|) \, \cos(k_{F\tau} |x|)
 \frac{e^{-\frac{|\alpha~\cos(\theta')|  \, \Delta  }{v_{F,sc} } }}{| \cos(\theta')|}\, \nn
&\hspace*{4.5cm} \times K_0\left( \frac{ |\alpha~ \text{sin}(\theta')| \Delta   } {u} \right) \int dX \,dT \cos\{2 \phi_1(X,T)\}  \nn
 =& \frac{ m_e \,\Delta \,\tilde t^2_{ext} \,\xi^2 \, L}{ \pi^2 \, \alpha} 
 \int_0^{\pi/2}   d\theta' \,
\sin (k_{F,sc}\, |x|) \, \cos(k_{F\tau} |x|) \frac{e^{-\frac{|\alpha~\text{cos}(\theta')|  \, \Delta  }{v_{F,sc} } }}{| \text{cos}(\theta')|}\, 
 K_0\left( \frac{ |\alpha~ \text{sin}(\theta')| \Delta   } {u} \right) \nn
&\hspace*{3cm}\times  \int \frac{u}{\pi \, \alpha^{2}} dX \,dT \cos\{2 \phi_1(X,T)\} \frac{d\alpha}{ \alpha}.
\end{align}

\begin{figure}
        \centering
                \includegraphics[width=0.5 \textwidth]{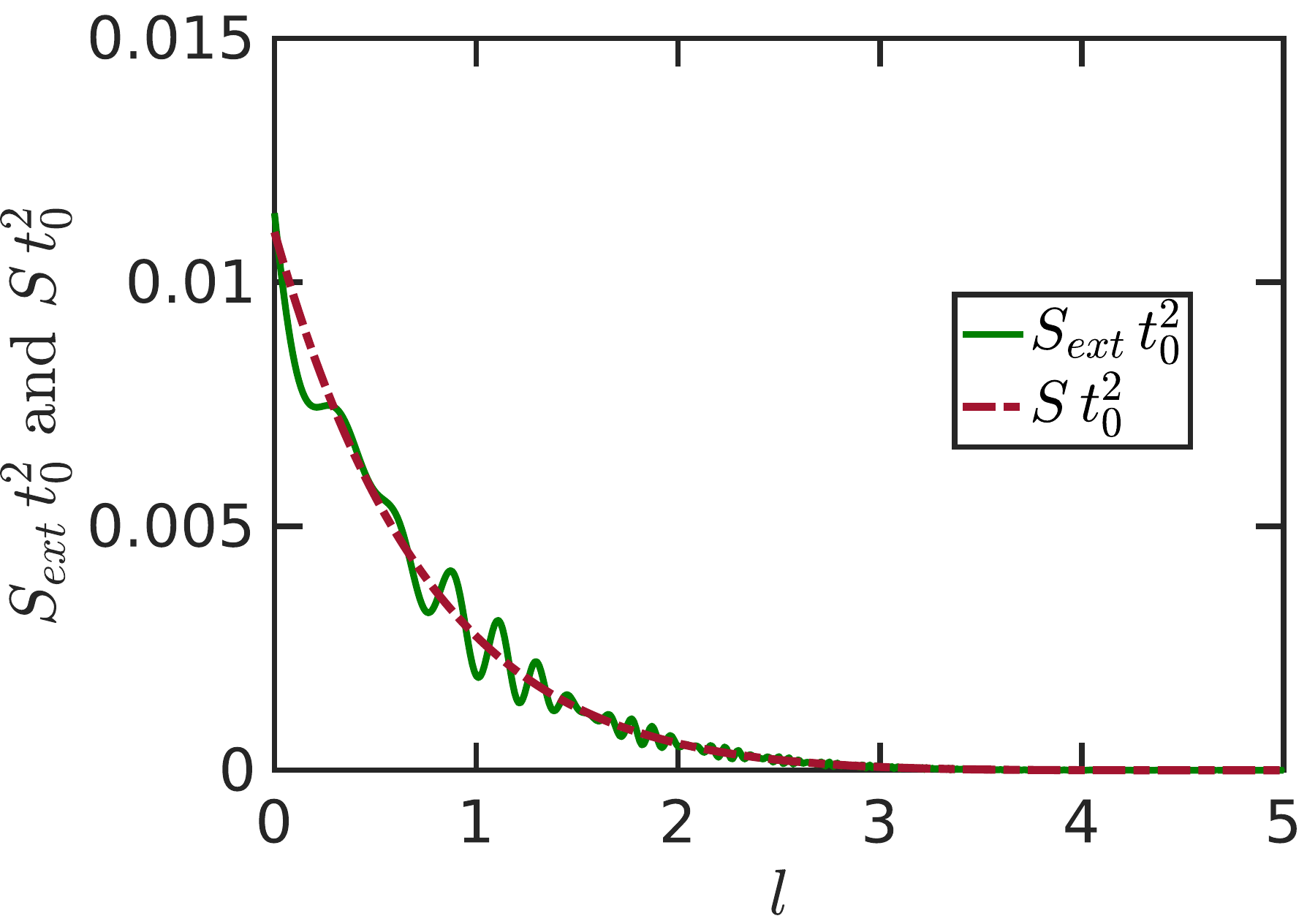} 
                \caption{ The comparison between two source terms  $S_{ext} t_0^2$ (green) and  $S t_0^2$ (red) represented as a function of the RG flow parameter $l$, which are calculated by using Eqs. (\ref{Sext}) and (\ref{S_ext}), respectively. Disregarding small oscillations, the agreement between two terms is fairly good.
        The parameter values are fixed to $t_0=\tilde t_{int}(0)=\tilde t_{ext}(0)=3.8 \times 10^{-5}$, $\Delta=0.35\, \text{meV}$,
$u=10^4~ \text{m/s}$,
$v_{F,sc}= 10^6~ \text{m/s}$, $\alpha_0=1 \,\text{nm}$,  $d=15 \,\alpha_0$, and $\alpha_{sc}=1/k_{F,sc}=1 \buildrel _\circ \over {\mathrm{A}}$. We use $k_{F\tau}=1/\alpha_0$, which is much higher than the realistic value of $k_{F\tau}$, to capture the maximum effect from it on $S_{ext}$, however, for realistic values of $k_{F\tau}$, the functional form of $S_{ext}$ comes out to be similar to $S_{int}$ plotted in Fig. \ref{fig_S_ap}.}
\label{fig_S_3_ap}
\end{figure}

By using $dl= \frac{d\alpha}{ \alpha} $ and $\xi= v_{F,sc}/ \Delta$, we arrive at the RG flow equation for $\tilde \Delta_\tau^{ext}$ in the form
\begin{align}
&\frac{d\tilde \Delta_\tau^{ext}}{dl}=\frac{ m_e \,v_{F,sc}^2 \,\tilde t^2_{ext} \, L}{\pi^2\,\Delta \, \alpha} \int_0^{\pi/2}   d\theta' 
\sin (k_{F,sc}\, |x|) \, \cos(k_{F\tau} |x|) \frac{e^{-\frac{|\alpha~\text{cos}(\theta')|  \, \Delta  }{v_{F,sc} } }}{| \text{cos}(\theta')|}\, 
 K_0\left( \frac{ |\alpha~ \text{sin}(\theta')| \Delta   } {u} \right) =S_{ext}\,\tilde t^2_{ext} \,,\\
&S_{ext} =\frac{ m_e v_{F,sc}^2  L}{2\,\pi^2\,\Delta \, \alpha}\,
 \int_0^{\pi/2}   d\theta' \,
\big[\sin \big\{(k_{F,sc}+k_{F\tau}) |x|\big\} + \sin \big\{(k_{F,sc}-k_{F\tau}) |x|\big\} \big] \frac{e^{-\frac{|\alpha~\text{cos}(\theta')|  \Delta  }{v_{F,sc} } }}{| \text{cos}(\theta')|}\, 
 K_0\left( \frac{ |\alpha~ \text{sin}(\theta')| \Delta   } {u} \right).
 \label{Sext}
\end{align}
We again split the integral into two parts $\int_0^{\pi/2}=\int_{\theta'_{c\pm}}^{\pi/2}+ \, \text{small contributions}$, where $\theta'_{c\pm}= \cos^{-1}(\frac{\pi}{2\,(k_{F,sc}\pm \,k_{F\tau})\,\alpha})=\pi/2 - \sin^{-1}(\frac{\pi}{2\,(k_{F,sc}\pm \, k_{F\tau})\,\alpha})=\pi/2\, [1 - \frac{1}{(k_{F,sc}\pm \, k_{F\tau})~\alpha}]$. We use $\int_{\pi/2}^{\pi/2-\delta\theta'} f(\theta') d\theta'= -f(\pi/2) \, \delta\theta'$ and approximate the integral in $S_{ext}$  as
 \begin{align}
 S&=\frac{ m_e \,v_{F,sc}^2 \, L}{ 2\,\pi^2 \, \Delta \, \alpha} 
 \Bigg[ \int_{\pi/2 - \frac{\pi}{2\,(k_{F,sc}+ \,k_{F\tau}) \,\alpha}}^{\pi/2}   d\theta' \,
\sin\left\{  (k_{F,sc}+k_{F\tau})\, | \alpha~\text{cos}(\theta') | \right\}
 \frac{e^{-\frac{|\alpha~\text{cos}(\theta')|  \, \Delta  }{v_{F,sc} } }}{| \text{cos}(\theta')|}\, 
 K_0\left( \frac{ |\alpha~ \text{sin}(\theta')| \Delta   } {u} \right)\nn
 &\hspace*{2.5cm}+ \int_{\pi/2 - \frac{\pi}{2\,(k_{F,sc}- \,k_{F\tau}) \,\alpha}}^{\pi/2}   d\theta' \,
\sin\left\{  (k_{F,sc}-k_{F\tau})\, | \alpha~\text{cos}(\theta') | \right\}
 \frac{e^{-\frac{|\alpha~\text{cos}(\theta')|  \, \Delta  }{v_{F,sc} } }}{| \text{cos}(\theta')|}\, 
 K_0\left( \frac{ |\alpha~ \text{sin}(\theta')| \Delta   } {u} \right)\Bigg]\nn
 &=\frac{ m_e \,v_{F,sc}^2  \, L}{  2\, \pi\,\Delta \, \alpha} \,
 K_0\left( \frac{ \alpha\, \Delta   } {u} \right).
 \label{S_ext}
 \end{align}
At the last step, we restore $\hbar$ and rewrite the final expression  as
$ S=\frac{ m_e \,v_{F,sc}^2  \, L}{ 2\, \pi\, \Delta \, \alpha} \,
 K_0\left( \frac{ \alpha\, \Delta   } {\hbar \,u} \right)$. In Fig. \ref{fig_S_3_ap}, we show that the approximate value ($S$) matches quite well with the exact one,  $S_{ext}$. Thus, for simplicity we can use $S$ when solving the RG flow equations numerically. 
 
At the last step, we collect all contributions coming from the tunneling Hamiltonian. The operator proportional to $\tilde t_{int}$ contains either $R_{1 \bar 1}$ or $L_{1 1}$ [$R_{\bar 1 \bar 1}$ or $L_{\bar 1 1}$] which in bosonic form are written as $e^{i(\phi_3-\theta_3)}$ or $e^{i(\phi_3+\theta_3)}$ [$e^{i(\phi_4-\theta_4)}$ or $e^{i(\phi_4+\theta_4)}$]. For identical interactions in NWs, $K_3=K_4$, the scaling dimension of $\tilde t_{int}$  is $\frac{K_3+1/K_3}{4}$, see Eq. (\ref{dc}). Similarly, the scaling dimension for $\tilde t_{ext}$ is $\frac{K_1+1/K_1}{4}$. For calculating the RG flow equations of the  remaining parameters $\tilde\Delta_{\tau}^{ext}, \tilde\Delta_{\tau}^{int}, \tilde\Delta_c, K_1$, and $K_3$, we follow the same procedure as in Appendix \ref{arg} and eventually obtain the RG equations given in Eq. (\ref{rg4}) of the main text.  Notably, in Eq. (\ref{rg4}) we do not include direct contributions from the tunneling terms to the renormalization of the LL parameters $K_1,...,K_4$ as they give rise to higher order terms which have negligible effect. Indeed, going to 4th order  in $t_{int/ext}$ in Eq.~(\ref{Zin}) and using  OPE we find that 
the resulting renormalization of the kinetic terms, Eq.~(\ref{kineticterms}), becomes proportional to $(S\tilde t_{int/ext}^2)^2$.
Such terms, however, vanish quickly under the RG flow, see Figs.~\ref{sfunc} and \ref{fig_S_2} in the main text. Thus, compared to the rapidly growing proximity gaps $\tilde\Delta_{\tau}^{ext}, \tilde\Delta_{\tau}^{int}$, and $ \tilde\Delta_c$, see  Fig.~\ref{fig_S_2}, we can safely neglect such direct contributions to $K_1,...,K_4$ in the RG equations (\ref{rg4}).

\twocolumngrid


\begin{thebibliography}{99}
\bibitem{footnote2} Note that we keep only the lowest order in the tunneling amplitude when we derive the RG equations. In the non-interacting case, this corresponds to the weak-coupling regime where all proximity gaps (being proportional to $t^2_{int/ext}$) are  smaller than the gap $\Delta$ of the bulk superconductor~\cite{Reeg_a}. Thus, for self-consistency we have to stop the RG flow whenever any of the proximity gaps reaches this upper limit $\Delta$.
\end{thebibliography}
\end{document}